\documentclass[12pt]{article}
\usepackage{amsmath,amssymb,amsfonts,amsthm}
\usepackage[dvips]{graphics}
\usepackage[dvips]{pstcol}
\usepackage{eucal}
\usepackage{pstricks,pst-plot,pst-node}
\newpsobject{showgrid}{psgrid}{subgriddiv=1,griddots=5,gridlabels=0pt}

\definecolor{blr}{rgb}{0.668, 0.0625, 0.9961}

\newcommand{\bl}{\textcolor{blue}}

\newcommand{\cM}{\mathcal{M}}

\newcommand{\C}{\mathbb{C}}
\newcommand{\bP}{\mathbf{P}}
\newcommand{\RP}{\mathbb{R}\mathbf{P}^2}

\newcommand{\Z}{\mathbb{Z}}
\newcommand{\R}{\mathbb{R}}
\newcommand{\cE}{\mathcal{E}}
\newcommand{\hs}{h^\star}

\newcommand{\bx}{\mathbf{x}}
\newcommand{\old}[1]{}
\newcommand{\Ronk}{\mathcal R}

\DeclareMathOperator{\const}{const}
\DeclareMathOperator{\dv}{div}
\DeclareMathOperator{\Area}{Area}
\DeclareMathOperator{\LogArea}{LogArea}
\DeclareMathOperator{\Ex}{\mathcal{E}xp}

\newtheorem{Theorem}{Theorem}
\newtheorem{Lemma}{Lemma}
\newtheorem{Corollary}[Lemma]{Corollary}
\newtheorem{Proposition}[Lemma]{Proposition}

\theoremstyle{definition}

\begin{document}
\title{Limit shapes and the complex Burgers equation}
\author{Richard Kenyon and Andrei Okounkov}
\date{} \maketitle

\tableofcontents

\section{Introduction}

\subsection{Overview}

In this paper we study surfaces in $\R^3$ that arise
as limit shapes in random surface
models related to planar dimers.
These limit shapes are
\emph{surface tension minimizers}, that is, they minimize
a functional of the form $\int \sigma(\nabla h) \, dx \, dy$ 
among all Lipschitz functions $h$ taking given values on the 
boundary of the domain. 
The surface tension $\sigma$ has singularities and is not
strictly convex, which leads to formation of
\emph{facets} and \emph{edges} in the limit shapes.

We find a change of variables that reduces the
Euler-Lagrange equation for the variational
problem to the complex inviscid Burgers equation (complex Hopf equation).
The equation can thus be solved in terms of an
arbitrary holomorphic function, which is somewhat
similar in spirit to Weierstrass parametrization of
minimal surfaces. We further show that for a
natural dense set of boundary conditions,
the holomorphic function in question is, in
fact, \emph{algebraic}. The tools of algebraic
geometry can thus be brought in to study the
minimizers and, especially, the formation of
their singularities. This is illustrated by
several explicitly computed examples.

\subsection{Random surface models}

The general class of random surface models to which
our results apply are the \emph{height functions} of
\emph{dimer models} on periodic weighted planar
bipartite graphs, see \cite{KOS} for an introduction.
For our purposes, the following simplest
model in this class is already very interesting.

\begin{figure}[!htbp]
  \centering
  \scalebox{0.6}{\includegraphics*[0,185][530,790]{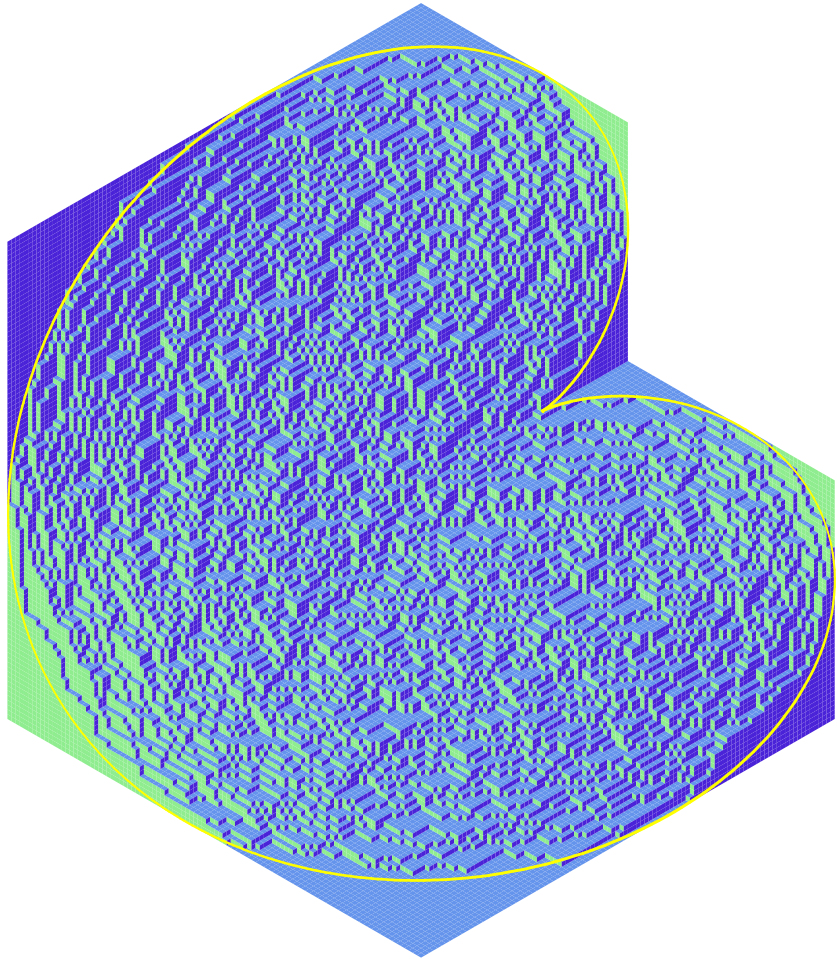}}
  \caption{A limit shape simulation.
The ``frozen boundary'' (yellow), which is asymptotically the boundary
of the facets, is a cardioid.}
  \label{fcard}
\end{figure}

Consider continuous surfaces in $\R^3$ glued out of
unit squares in coordinate directions, that is, out of 
parallel translates of the faces of the unit
cube. We assume that the surfaces are \emph{monotone},
that is, project 1-to-1 in the (1,1,1) direction, and
span a given polygonal contour $C$, see Figure \ref{fcard}
for an illustration. We call such surfaces \emph{stepped surfaces}.
Uniform measure
on all such stepped surfaces is an obvious 2-dimensional
generalization of the simple random walk on $\Z$.
It is well-known and obvious from Figure \ref{fcard} that
stepped surfaces spanning a given contour $C$ are
in a bijection with tilings of the region $\Omega$ enclosed
by the $(1,1,1)$-projection of $C$
with three kinds of rhombi, known as \emph{lozenges}. They are also
in bijection with dimer coverings---that is, perfect matchings---of the hexagonal graph.

Let $C_n$ be a sequence of boundary contours each of which
can be spanned by at least one stepped surface; we will
call such contours \emph{tilable}. Suppose
that $n^{-1}\, C_n$ converges to a given curve
$C\subset \R^3$. In this case $C$ can be spanned by
a Lipschitz surface whose normal (which exists almost everywhere)
points into the positive
octant (any limit of stepped surfaces of the $C_n$ is such a surface); we
will call such contours $C$ \emph{feasible}.
It is well-known, see e.g.\ \cite{Fournier} that feasibility is equivalent
to being a limit of tilable contours.

It is natural to consider the weak
limit of the corresponding uniform measures on
stepped surfaces of $C_n$, scaled by $n^{-1}$.
A form of the law of large numbers, proved in \cite{CKP},
says that this limit depends only on the contour $C$ and is, in
fact, a $\delta$-measure on a single Lipschitz
surface spanning $C$. This surface, known
as the \emph{limit shape}, can, for
example, be interpreted as the macroscopic shape
of an interface (e.g.\ crystal surface) arising from its microscopic
description and boundary conditions.

A question of obvious interest is to describe
this limit shape quantitatively and qualitatively.
In particular, an important phenomenon, observed in
nature and reproduced in this simple model, is
formation of facets, that is, flat pieces,
in the limit shape.

\subsection{Variational principle}

\begin{figure}[!htbp]
  \centering
  \scalebox{0.9}{\includegraphics{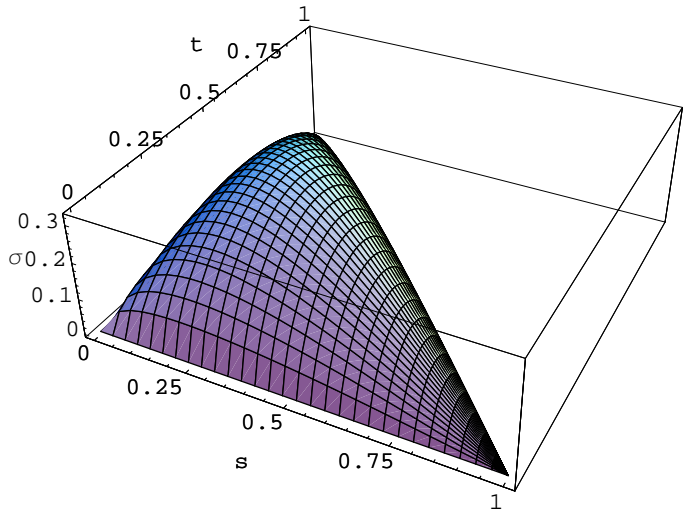}}
  \caption{Graph of $-\sigma$ for the line $P(z,w)=z+w-1$.
  \label{fig_surf_tension}}
\end{figure}

The following variational characterization of the limit
shape was proved in \cite{CKP}. The limit shape may be
parameterized as the graph of a function
$$
x_3 = h(x_3-x_1,x_3-x_2)
$$
where $h$ is a Lipschitz function with gradient
$\nabla h$ (which is defined almost everywhere by Rademacher's
theorem) in the triangle
\begin{equation}
  \label{triang}
  \Delta=\textup{conv}\, \big\{(0,0),(0,1),(1,0)\big\}\,.
\end{equation}
Let $\Omega$ be the plane region enclosed by the
projection of the boundary curve $C$ along the
$(1,1,1)$-direction. We will use
$$
(x,y) = (x_3-x_1,x_3-x_2)
$$
as coordinates on $\Omega$.
The limit shape height function $h$ is the unique minimizer of the functional
\begin{equation}
  \label{variat}
    \text{ENT}(h) = \int_\Omega \sigma(\nabla h(x,y)) \, dx \, dy \to \min
\end{equation}
among all Lipschitz functions with the given boundary conditions
(that is, for which the graph of $h$, restricted to $\partial\Omega$, is $C$)
and gradient in $\Delta$. Here $\sigma$ is the \emph{surface
tension}. It has an explicit form which, in the language of \cite{KOS},
is the Legendre
dual of the \emph{Ronkin function} of the simplest plane curve
$$
z+w = 1 \,.
$$
We recall that for a plane curve $P(z,w)=0$, its Ronkin function \cite{Mikh}
is defined by
$$
\Ronk(x,y) = \frac1{(2\pi i)^2} \iint_{
  \substack{|z|=e^x\\|w|=e^y}} \log \big|P(z,w)\big| \, \frac{dz}{z} \,
\frac{dw}{w}\,.
$$
The gradient $\nabla\Ronk$ always takes values in the Newton polygon
$\Delta(P)$ of the polynomial $P$, so $\Delta(P)$ is naturally
the domain of the Legendre
transform $\sigma=\Ronk^\vee$. For the straight line as above,
the Newton polygon is evidently the triangle \eqref{triang}. See Figure
\ref{fig_surf_tension}.

\subsection{General planar graphs}
More generally, to an arbitrary planar periodic bipartite edge-weighted
graph (with periodic edge weights)
one associates a real plane curve $P(z,w)=0$, its \emph{spectral
curve}. Limit shapes for the height function in the corresponding
dimer problem satisfy the same variational principle as above
with $\sigma=\Ronk^\vee$,
where $\Ronk$ is the Ronkin function of $P(z,w)$, see \cite{KOS}.
The spectral curve $P$ of the dimer model
is always a so-called \emph{Harnack curve}. This property, also
known as
\emph{maximality},  is crucial for the constructions in this paper.
Harnack curves form a very special class of real plane curves,
see \cite{KO}. Many probabilistic implications of maximality are
discussed in \cite{KOS}.

Maximality, in particular, implies that the surface tension $\sigma$
is analytic and strictly convex in the interior $\Delta(P)^\circ$
of $\Delta(P)$,
with the exception of lattice points, that is, points of $\Z^2$.
At a lattice point, generically, $\sigma$ has a conical singularity, often
referred to as a \emph{cusp} in physics literature. For example, at
the point $(0,0)$ in Figure \ref{fig_surf_tension}, $\sigma$ has
the form
$$
\sigma(s,t) = t \ln\frac{t}{t+s} + s\ln\frac{s}{t+s} +o(s,t)\,,
$$
in which the reader will recognize the Shannon entropy. More generally,
at a lattice point, the graph of $\sigma$ is approximated by
a cone over a curve which is the
polar dual of the curve bounding the corresponding facet of the Ronkin function. Such
singularities lead to facet
formation in the minimizer. They disappear only
if the spectral curve has a nodal singularity, which is a
codimension two condition on the weights in the dimer model.

At the
boundary of the Newton polygon, $\sigma$ is piecewise-linear and,
in particular, not strictly convex. This loss of strict convexity
leads to edges in the limit shape. Both facets and edges are clearly
visible in the simulation in Figure \ref{fcard}.

These features of the surface tension make the analytic
study of the variational problem \eqref{variat}
challenging and interesting.

\subsection{Complex Burgers equation}

We say that a point $(x_0,y_0)\in\Omega$ is in the \emph{liquid
region} if the minimizer $h(x,y)$ is $C^1$ in a
neighborhood of $(x_0,y_0)$ and $\nabla h (x_0,y_0)$ is either
in $\Delta(P)^\circ \setminus \Z^2$ or at a point (in $\Z^2$)
corresponding to a nodal singularity of $P$.
This terminology is motivated by the frozen/liquid/gaseous
classification of phases of a dimer model \cite{KOS}.

Since $\sigma$ is analytic and strictly convex on the liquid
region, Morrey's theorem \cite{Morrey} implies that $h$ is analytic
there and satisfies the Euler-Lagrange
equation $\dv \nabla\sigma \circ \nabla h = 0$.
A more general equation
\begin{equation}
  \label{EL}
  \dv \nabla\sigma \circ \nabla h = c\,,
\end{equation}
where $c$ is a real constant, arises in the \emph{volume-constrained} minimization
problem
\begin{equation}
  \label{varp}
 \cE(h) =\int_\Omega \sigma(\nabla h(x,y)) \, dx \, dy  + c \int_\Omega h(x,y) \, dx \, dy \to \min \,,
\end{equation}
in which $c$ is the Lagrange multiplier
for the constraint
$$
\int_\Omega h(x,y) \, dx \,dy  = \const
$$
that forces the limit shape to enclose a given volume
(or more precisely, to contain a given volume under its graph).

Our first result is the following reduction of the
Euler-Lagrange equation \eqref{EL} to a first-order
quasilinear equation for a complex valued function $z(x,y)$.

\begin{Theorem}\label{t1} In the liquid region, we have
  \begin{equation}
    \label{arg}
    \nabla h = \frac{1}{\pi} (\arg w, - \arg z)
\end{equation}
where the functions $z$ and $w$ solve the differential equation
\begin{equation}
  \label{Burg}
  \frac{z_x}{z} + \frac{w_y}{w} = c
\end{equation}
and the algebraic equation $P(z,w)=0$ \,.
\end{Theorem}

Note that in \eqref{arg} the argument of a complex number is
a multivalued function and the statement is that there exists a
branch of the argument for which equality \eqref{arg} is
satisfied.

For example, in the lozenge tiling case
\begin{equation}
  \label{loz}
  P(z,w)=z+w-1
\end{equation}
and without the volume
constraint, equation \eqref{Burg} becomes
$$
\frac{z_x}{z_y} = \frac{z}{1-z}
$$
which becomes the standard complex inviscid Burgers equation
\begin{equation}
  \label{burg}
  u_x = u u_y
\end{equation}
upon the substitution $u=z/(1-z)$. In general,
equation \eqref{Burg}, being a first-order
quasilinear equation for $z(x,y)$, has properties
parallel to those of \eqref{burg} and, in particular, it
can be solved by complex characteristics as follows:

\begin{Corollary}\label{c1} For $c\ne 0$
there exists an analytic function $Q$ of two variables
such that
$$
Q(e^{-cx} z, e^{-cy} w) =0
$$
in the liquid region.
When $c=0$ there is an
analytic function $Q_0$ of two variables such that
\begin{equation}\label{Q when c=0}
Q_0(z,w)=xzP_z + ywP_w.
\end{equation}
\end{Corollary}

In other words (for $c\ne0$), as functions of $x$ and $y$, $z$ and $w$
are found by solving the system
\begin{equation}
  \label{PQ}
  \begin{cases}
     P(z,w)=0\,, \\
  Q(e^{-cx} z, e^{-cy} w) =0\,.
  \end{cases}
 \end{equation}
A general solution to the Euler-Lagrange equation \eqref{EL}
is thus parameterized by an analytic curve $Q=0$ in the
plane.

Below we define a dense set of boundary conditions for which
$Q$ is in fact algebraic. In this case \eqref{PQ} is a system of
algebraic equations and, hence,
$h$ is an integral of an algebraic function of $e^{-cx}$ and
$e^{-cy}$.

\subsection{Frozen boundary}

The liquid region may extend all the way to the boundary of
$\Omega$, but it may happen that it ends at a facet,
that is, a domain where the limit shape height
function $h$ is a linear function with integral gradient, as in Figure \ref{fcard}.
We call the boundary between the liquid region and the facets
the \emph{frozen boundary}.

There are two kinds of facets:
the ones with gradient $\nabla h$ in the
interior of the Newton polygon $\Delta(P)$ and the ones
with $\nabla h$ on the boundary of $\Delta(P)$. While the
corresponding microscopic properties of random
surfaces are, in a sense, diametrically opposite
(they correspond to gaseous and frozen phases,
respectively, in the classification of \cite{KOS}), the
only difference relevant for this paper is that only
gaseous facets can occur in the interior of the domain. 
We will call such floating facets \emph{bubbles}.

In terms of \eqref{arg}, a point $(z,w)$ of the frozen boundary
is a real point of the spectral curve
$P(z,w)=0$. A curve's worth of such real solutions of \eqref{PQ}
is possible only if $Q$ is real. In this case, at
the frozen boundary, the solutions $(z,w)$ and
$(\bar z,\bar w)$  of the system \eqref{PQ} become identical. That is,
the frozen boundary is a \emph{shock} for the equation \eqref{Burg}.
In contrast to the usual real-valued Burgers equation,
a shock is a special event in the complex case
as complex characteristics tend to miss each other.

Given two analytic functions $P$ and $Q$, the locus
where the system \eqref{PQ} has a double root is
itself an analytic curve in the $(x,y)$-plane.
In the case when $P$ and $Q$ are algebraic,
this discriminant curve is of the
form
\begin{equation}
  \label{Rcurv}
   \tilde R(e^{-cx},e^{-cy})=0\,,
\end{equation}
where $\tilde R$ is polynomial
which can be written down effectively in terms of
resultants.

The frozen boundary $R$ is a subset
(in principle, proper) of the discriminant
curve \eqref{Rcurv}. A more invariant way of saying the
same thing is that the map
$$
\Ex: (x_1,x_2,x_3) \mapsto \left(e^{cx_1}: e^{cx_2} : e^{cx_3}\right)
\in \R\bP^2
$$
from $\R^3$ to the real projective plane
takes the frozen boundary to a real algebraic curve.
As $c\to 0$ the map $\Ex$, suitably rescaled, becomes the
projection in the $(1,1,1)$ direction. Monotonicity
insures that our random surfaces are mapped 1:1 by
$\Ex$. Indeed,
if $(x_1,x_2,x_3)$ is on the surface then $(x_1+t,x_2+t,x_3+t)$ is on the
surface only for $t=0$.

At a generic point, the frozen boundary is smooth
and \eqref{PQ} has exactly one real double root.
The functions $z(x,y)$ and $w(x,y)$ have a square-root
singularity at the frozen boundary and, hence,
the arguments of $z$ and $w$ grow like the square root
of the distance as one moves from the frozen boundary
inside the liquid region. Integrating \eqref{arg},
we conclude that the limit shape height
function has an $x^{3/2}$ singularity at the
frozen boundary. Thus one recovers the
well-known Pokrovsky-Talapov law \cite{PT} in this
situation.

At special points of the frozen boundary, triple
solutions of \eqref{PQ} occur. At such points,
the frozen boundary has a cusp singularity. One
such point can be seen in Figure \ref{fcard}.

\subsection{Algebraic solutions}

Any boundary contour can be approximated by
piecewise linear contours and even by
piecewise linear contours $C$ with segments in
coordinate directions. For such contours the
function $Q$ is algebraic (of degree
growing with the number of segments). In this paper
we will prove this in the simplest case when
$P(z,w)=z+w-1$ and the boundary $C$ is connected.
The precise result that we
will prove is the following

\begin{Theorem}\label{t2}
Suppose $P(z,w)=z+w-1$ and the boundary contour $C$ is
feasible, connected, and
polygonal with $3d$ sides in coordinate directions
(cyclically repeated).
Then $Q$ is an algebraic curve of degree
$d$ and genus zero.
\end{Theorem}

The corresponding result for cases in which some of the edge lengths
are allowed to be zero can be obtained as limits,
letting the edge lengths shrink.
In these cases $Q$ may have smaller degree.

\subsection{Reconstruction of limit shape from the Burgers equation}\label{s_alg}
\label{s_recons} 

The frozen boundary $R$ is a genus zero
curve inscribed in the polygon $\Ex(C)$. The
reconstruction of the limit shape
from $R$ has the following elementary
geometric interpretation.

Let $\bx=(x_1,x_2,x_3)$ be a point on the limit shape
in the liquid region. Then $\Ex(\bx)$ lies inside
the frozen boundary $R$. In the course of proving Theorem \ref{t2},
we will show that there is a unique, up to conjugation,
complex tangent to $R$ through the point $\Ex(\bx)$.
This is illustrated in the Figure \ref{f_card_tan}.
There are 3 tangent lines to a cardioid through a general
point in the plane (this number of tangents
is the degree of the dual
curve and is classically known as the \emph{class} of a curve).
Through a point outside the cardioid, all 3 of these lines
are real. Through a point inside the cardioid, only one real
tangent exists, the other two tangents are complex.

\psset{unit=1 cm}
\begin{figure}[!htbp]
  \centering
   \begin{pspicture}(0,0)(5,4)
\rput[c](2.5,2){\scalebox{0.33}{\includegraphics{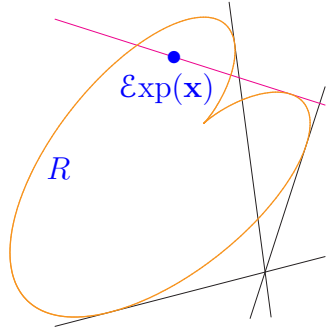}}}
\rput[c](2.6,3.42){$\bl{\bullet}$}
\rput[c](2.5,3){$\bl{\Ex(\bx)}$}
\rput[tl](.9,2.1){$\bl{R}$}
\end{pspicture}
 \caption{Tangent lines to a cardioid}
  \label{f_card_tan}
\end{figure}

\begin{figure}[!htbp]
  \centering
  \begin{pspicture}(-0.5,0)(4,3)
\psline[linecolor=green,linewidth=1.5pt]{->}(0,0)(1,3)
\psline[linecolor=green,linewidth=1.5pt]{->}(1,3)(4,1)
\psline[linecolor=green,linewidth=1.5pt]{->}(4,1)(0,0)
\psarc[linecolor=red](0,0){1}{14.04}{71.56}
\psarc[linecolor=red](1,3){1}{-108.43}{-33.69}
\psarc[linecolor=red](4,1){1}{146.31}{194.04}
\rput[c](0.1,1.8){\bl{$a_1 e^{c x_1}$}}
\rput[c](2.7,2.4){\bl{$a_2 e^{c x_2}$}}
\rput[c](2.1,0.2){\bl{$a_3 e^{c x_3}$}}
\rput[bl](0.74,0.71){\bl{$\alpha_2$}}
\rput[tl](1.3,2){\bl{$\alpha_3$}}
\rput[rb](3,1.1){\bl{$\alpha_1$}}
\end{pspicture}
\caption{A complex tangent defines a triangle}
  \label{f_triang}
\end{figure}
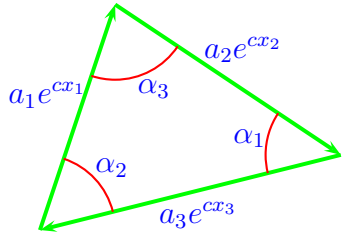

Let the complex tangent through $\Ex(\bx)$ be given by
the equation
$$
a_1 z_1 + a_2 z_2 + a_3 z_3 = 0
$$
in homogeneous coordinates $(z_1:z_2:z_3)$ on the
projective plane. Since the point $\Ex(\bx)$ lies on
this line, this gives us a triangle in the complex plane
illustrated in Figure \ref{f_triang}.

The numbers $(a_1,a_2,a_3)$ are unique up to
common complex factor and complex conjugation. If
follows that the similarity class of the triangle in
Figure \ref{f_triang} is well defined, in particular,
its angles $\alpha_1,\alpha_2,\alpha_3$ are well
defined. Formula \eqref{arg} simply says that
$$
(\alpha_1,\alpha_2,\alpha_3) \in \R_{>0}^3
$$
is the normal to the limit shape at the point $\bx$.

As the point $\bx$ approaches the frozen boundary,
the complex tangent approaches the corresponding
real tangent, the triangle degenerates, and the
normal starts pointing in one of the coordinate
directions.

Higher genus frozen boundaries occur for
multiply connected liquid regions. The holes
may be created manually  by considering
surfaces spanning a disconnected
boundary contour or they may appear as bubbles for spectral
curves $P$ of genus $\ge 1$. Examples of
such situations will be considered in Sections \ref{nonsimplyconnsec} and
\ref{bubbleex}. In the multiply-connected case, the 
Burgers equation needs to be supplemented by 
period conditions, see Sections \ref{is_b}.

\subsection{Burgers equation in random matrix theory}

In the case when $\Omega$ is an infinite vertical strip, the
random surface model studied here can be conveniently restated
in terms of nonintersecting lattice paths connecting given
points on $\partial\Omega$. The continuous limit of such
nonintersecting path models is well-known to be described by
the Hermitian $2$-matrix model (in other words, by a random
walk on the space of Hermitian matrices with boundary
conditions in given conjugacy classes). In physics literature,
complex Burgers equation first appeared in this context in
the work of Matytsin \cite{Mat}. Its rigorous mathematical study
appears in the work of A.~Guionnet \cite{Gui}. Generalizations
of these equations to interacting particle systems are
discussed in \cite{Aba}.

\subsection{Acknowledgments}

The paper was started during the authors' visit to
Institut Henri Poincar\'e in the Spring of 2003.
The work was continued while R.K.\ was visiting
Princeton University. We thank these
institutions for their hospitality.

The work of R.K.\ was partially supported by CNRS and NSERC.
A.O.\ was partially supported by the NSF and a fellowship
from the Packard Foundation.

We benefited from discussions with A.~Abanov, A.~Guionnet,
N.~Reshetikhin, I.~Rodnianski, S.~Sheffield, and S.~Smirnov.

\section{Complex Burgers equation}

\subsection{Proof of Theorem \ref{t1}}

By the basic properties of the Ronkin function, see \cite{Mikh}, the
function
$$
\nabla\sigma = (\nabla \Ronk)^{-1} : \R^2 \to \R^2
$$
takes values in the \emph{amoeba} of the spectral curve. By definition,
this means that for any $(x,y)$ in the liquid region we can find
a point $(z,w)$ on the spectral curve
satisfying
\begin{equation}
  \label{lift}
  \big[\nabla\sigma \circ \nabla h \big]
(x,y)= (\log |z|,\log |w|) \,.
\end{equation}
By our hypothesis, near $(x,y)$ the LHS of \eqref{lift}
lies in the interior of the amoeba.
One of the several equivalent definitions of a Harnack curve
is that the map
\begin{equation}
  \label{am_map}
   (z,w) \mapsto (\log |z|,\log |w|)
\end{equation}
from the curve to its amoeba is 2-to-1 with
nonzero Jacobian over the amoeba's interior.
Therefore, the lift $(x,y)\mapsto (z,w)$ in \eqref{lift}
is unique up to complex conjugation, and
real analytic in a neighborhood of $(x_0,y_0)$.

Using \eqref{lift}, the Euler-Lagrange equation \eqref{EL} stated in
terms of $(z,w)$ becomes
$$
\Re \left(\frac{z_x}{z} + \frac{w_y}{w} \right) = c \,,
$$
This first order equation for the gradient $\nabla h$
needs to be supplemented by the usual consistency relation
for the mixed partials:
\begin{equation}
  \label{mixed}
   \left(-\frac{\partial}{\partial y}, \frac{\partial}{\partial x}
\right) \cdot \nabla h =
\left(-\frac{\partial}{\partial y}, \frac{\partial}{\partial x}
\right) \cdot \nabla \Ronk \, (\log |z|,\log |w|) = 0 \,.
\end{equation}
Again, it is a special property of the Harnack curve that
\begin{equation}
  \label{Ronkin_grad}
  \nabla \Ronk \,  (\log |z|,\log |w|) =
\frac1{\pi}\left(\mp \arg w, \pm \arg z\right)
\end{equation}
where on the right we mean a lift of $(\mp \arg w, \pm \arg z)$ and
where the choice of signs depends on the choice of
one of the preimages in \eqref{am_map}, see \cite{PR} and
also \cite{Mikh}. This turns the relation \eqref{mixed} into
the imaginary part of \eqref{Burg} (since $c\in\R$) and concludes the
proof.

\subsection{Proof of Corollary \ref{c1}}

Since $z$ and $w$ satisfy an algebraic equation $P(z,w)=0$,
their Jacobian vanishes
$$
J(z,w)=\det
\begin{pmatrix}
z_x & z_y \\
w_x & w_y
\end{pmatrix}
= 0
\,.
$$
We have
$$
J(e^{-cx}z,e^{-cy}w) - J(z,w) = c^2  z w - c z w_y - c z_x w = 0
$$
by \eqref{Burg} and hence $J(e^{-cx}z,e^{-cy}w)=0$ identically. The functions
$e^{-cx}z$ and $e^{-cy}w$ are real analytic in the liquid region and
the vanishing of the Jacobian implies a functional dependence of
the form
$$
Q(e^{-cx}z,e^{-cy}w)=0\,,
$$
where $Q$ is an analytic function (note that if $W=W(x,y),Z=Z(x,y)$ satisfy
$Z_xW_y=Z_yW_x$, then $\partial W/\partial\bar Z=
W_x/ Z_x-W_y/Z_y=0$,
so that $W$ is an analytic function of $Z$, and vice versa.)

\subsection{Complex structure on the liquid region}

Consider the function
$$
\zeta = e^{-cx} z
$$
which is defined on the liquid region. Differentiating $P(z,w)$
with respect to $y$ and using \eqref{Burg}
implies that
\begin{equation}
  \label{zeta_x}
  \frac{\zeta_x}{\zeta_y} = \frac{z P_z}{w P_w}
\end{equation}
when $P(z,w)=0$.
The rational function
$$
(z,w) \mapsto \gamma(z) = \frac{z P_z}{w P_w} \in \C\bP^1
$$
is known as the \emph{logarithmic Gau\ss\ map} of the
spectral curve $P(z,w)=0$.

By a result of Mikhalkin \cite{Mikh},
the preimage $\gamma^{-1}\left(\R\bP^1\right)$ is
precisely the critical locus of the map from the
curve to its amoeba. For a Harnack curve, therefore,
it coincides with the real locus. It follows that
on the liquid region $\gamma$ takes
values in upper or lower half-plane, depending on
the choice of lift of \eqref{lift}. With one
of the choices, the map
$$
(x,y) \to \zeta
$$
is orientation-preserving and we can use it
to pull back the complex structure.

Note that one could equally well take the function
$e^{-cy} w$ as defining the complex structure on the
liquid region. However, by Corollary \ref{c1} the
complex structure thus obtained is the same.
Yet another way to say this is that
$$
(x,y) \mapsto (e^{-cx}z,e^{-cy}w)
$$
maps the liquid region to the analytic curve $Q=0$
and the complex structure on the liquid region
is the unique one making this map holomorphic.

The orientation-preserving property of $\zeta(x,y)$
can be strengthened as follows

\begin{Proposition}\label{diffeo} For $c\ne 0$ and $(x,y)$ in the 
liquid region, the map
\begin{equation}
  \label{map_toQ}
     \pi_Q: (x,y) \mapsto (e^{-cx}z,e^{-cy}w) \in Q
\end{equation}
is an orientation preserving local diffeomorphism onto its 
image. In the case $P(z,w)=z+w-1$ the map is a diffeomorphism
onto its image.
\end{Proposition}

\begin{proof} 
Suppose that $\pi_Q$ maps two distinct 
points $(x_1,y_1)$ and $(x_2,y_2)$ to the 
same point
$$
(e^{-cx_1} z_1, e^{-cy_1}w_1) = (e^{-cx_2} z_2, e^{-cy_2}w_2) \in Q
$$
of $Q$, where $z_i=z(x_i,y_i)$. This means that on the 
curve $P$ we have a pair of distinct points 
$$
P\owns (z_1,w_1) \ne (z_2,w_2) \in P
$$
with the same arguments
$$
\arg (z_1,w_1)  = \arg (z_2,w_2) \,.
$$
However, $P$ is a Harnack curve and for a such a curve we have 
\eqref{Ronkin_grad}. By strict convexity of the Ronkin 
function, its
gradient maps both components of 
$P(\C)\setminus P(\R)$ one-to-one to the Newton polygon of $P$,
and, as a result, a point of $P$ is locally determined 
by the arguments. If $P=z+w-1$ (or even if $P$ has degree $\leq 2$), 
a point of $P$ in a component of $P(\C)\setminus P(\R)$ is determined
uniquely by its arguments.
\end{proof}

There are examples where $\pi_Q$ is a non-trivial covering map.

In the case $c=0$, we can use
$z$ or $w$ to define the complex structure.
This complex structure is
used in \cite{K.fluct} in the study
of the fluctuations of the random surfaces. Conjecturally,
the Gaussian correction to the limit shape is given
by the massless free field in the conformal structure
defined by $\zeta$.

\subsection{Genus of $Q$}

Suppose that the liquid region is surrounded by the 
frozen boundary, that is, suppose that $\Im z, \Im w \to 0$
at the boundary of the liquid region. The real parts 
$\Re z$ and $\Re w$ may or may not be continuous at the 
boundary of the liquid region.

\begin{Proposition}\label{genusQ}
If the liquid region is surrounded by the 
frozen boundary and the functions $z(x,y)$ and $w(x,y)$
are continuous up to the frozen boundary then 
the curve $Q$ is algebraic. If $P=z+w-1$ then 
$Q$ is of the same genus as 
the liquid region. 
\end{Proposition}

\begin{proof}
If $z,w$ are continuous then the map \eqref{map_toQ},
and its complex conjugate glue along the 
boundary of the liquid region to a map from 
the double of the liquid region to the curve $Q$.
This map is orientation preserving and unramified
by Proposition \ref{diffeo}, and so a covering map of $Q$.
It follows that the 
analytic curve $Q$ is compact. If $P=z+w-1$, the
map is a diffeomorphism and therefore $Q$ is of the same 
genus as the liquid region. 
\end{proof}

Conversely, if $Q$ is algebraic then the solutions
of \eqref{PQ} are continuous. However, only for 
some very special curves $Q$ will the solution 
of \eqref{PQ} satisfy the global 
requirements of being 
single-valued in some region and 
becoming real at the region's boundary.

By taking limits of algebraic solutions, one 
can construct solutions having, for example, 
accumulation points of cusps. At such a point 
$z$ is discontinuous. Namely, $z$ turns 
infinitely many times around $\R\bP^1$ as 
one approaches the accumulation point along the frozen 
boundary. 

It would be interesting to prove the continuity 
required in Proposition \ref{genusQ} directly from 
the variational problem and polygonality of 
the boundary conditions. In this paper, we take 
a different route and construct the algebraic 
curve $Q$ by a deformation argument. 

Note that Proposition \ref{genusQ} forces $Q$ to have 
the maximal possible number of real ovals (real components), namely 
genus plus one. Such real curves are known as 
\emph{$M$-curves}.

\subsection{Islands and bubbles}\label{is_b}

\begin{figure}[htbp]
 \begin{center}
  \scalebox{0.75}{\includegraphics{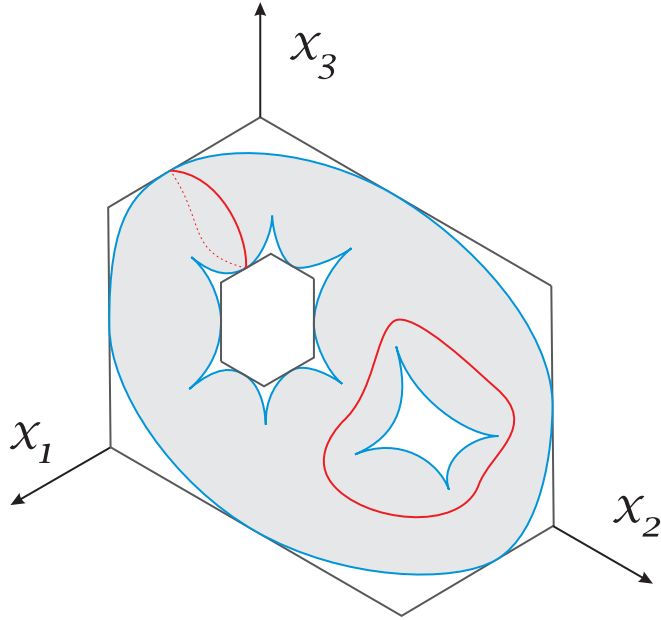}}
  \end{center}
 \caption{An island, a bubble, and their period conditions. 
\label{isl_bub}}
\end{figure}

Holes
in the liquid region may arise by considering
surfaces spanning a disconnected
boundary contour or they may appear as gaseous bubbles
(where the surface tension has a cusp).
We will call such holes \emph{islands} and \emph{bubbles},
respectively. A schematic representation of an 
island and a bubble can be seen in Figure \ref{isl_bub}

In principle, the solutions of a system like \eqref{PQ}
may pick up a monodromy around a nontrivial
loop in the liquid region. However,
in our case we require them to be
single-valued because they are determined by the
single-valued function $\nabla h$.

In the presence of holes, the Burgers equation 
\eqref{Burg} needs to be supplement by
integral constraints. This constraints can be phrased as 
imposing conditions on the periods of the curve $Q$.

\subsubsection{Areas of bubbles}\label{s_per}

In the case of a bubble inside the liquid region, it
is more convenient to use the integrated form of \eqref{EL}
which says that the flow of the vector field
$\nabla\sigma \circ \nabla h$
through any closed contour equals $c$ times the
area enclosed. 

Let a domain $B\subset\Omega$ be such 
that its boundary $\partial B$ is smooth and 
lies in the liquid region. 
Consider a variation of $h$ which is a smooth 
approximation to the characteristic function of $B$. 
This yields the real part of the equation 
\begin{equation}
  \label{circA}
  \int_{\partial B} \alpha  = c \, \Area(B) \,.
\end{equation}
where $\alpha$ is the following $1$-form
$$
\alpha = \log z \, dy - \log w \, dx \,. 
$$
By \eqref{arg}, the vanishing of the
imaginary part of \eqref{circA} means that the height
function $h$ is well-defined. Thus \eqref{circA} is
satisfied by any $B$ as above. 

Suppose that $(z,w)$ are continuous up to the frozen boundary.
On the frozen boundary, 
the points $(e^{-cx}z,e^{-cy}w)$ and $(z,w)$ are real 
points of $Q$ and $P$, respectively. This allows us to associate,
to each oval of $Q$, a corresponding oval of $P$. 
We will call an oval \emph{compact} if it does not 
intersect the coordinate axes. Note that an oval of $Q$ 
is compact if and only if the corresponding oval of $P$ is. 
Because $P$ is a Harnack curve, it has exactly one 
noncompact oval that intersects each
coordinate axis $\deg P$ times.

If $B$ is a bubble, a point on its boundary gives a point
$(z,w)\in P$ belonging to a compact oval of $P$. 
Integrating \eqref{circA} around the boundary of $B$,
the imaginary part of the left-hand side of \eqref{circA}
is then zero, since on the boundary of a bubble $z$ and $w$ are
real and of definite sign.
However the real part may not vanish and represents a 
nontrivial constraint of the curve $Q$. This constraint 
can be reformulated as follows. 

Given a compact oval $O$ of an plane curve, consider 
its \emph{logarithmic area} 
$$
\LogArea(O) = \int_O \log z \, d\log w \,,
$$
that is, the area enclosed by the image of $O$ under 
the amoeba map. 

\begin{Proposition}\label{log_arias}
The logarithmic areas of compact ovals of $Q$ are multiples of 
the logarithmic areas of the corresponding ovals of $P$. 
The multiple at a given oval is given by the winding number of the map from 
the corresponding bubble $B$ to the oval of $Q$.
\end{Proposition}

\begin{proof}
Let $O_P$ and $O_Q$ be the ovals of $P$ and $Q$ 
corresponding to a bubble $B$. Because these 
ovals are compact, the logarithms of $z$, $w$, 
$\zeta=e^{-cx} z$ and $\omega=e^{-cy} w$ are 
well-defined on $\partial B$. We have 
\begin{multline*}
\LogArea(O_Q)  = 
\int_{\partial B} \log \zeta \, d\log \omega  = \\ 
c^2\int_{\partial B} x \, dy-c\int_{\partial B} (y\,d\log z+x\, d\log w)+\int_{\partial B}\log z\,d\log w =\\
c^2\Area(B)- c\int_{\partial B} \alpha +m\LogArea(O_P)=m\LogArea(O_P)\,,
\end{multline*}
where $m$ is the winding number of $\partial B$ around $O_Q$.
\end{proof}

\subsubsection{Heights of islands}\label{s_high}

Different period conditions
need to be imposed if $\Omega$ has interior boundary as
in Figure \ref{isl_bub} or in the example considered in Section
\ref{nonsimplyconnsec}. In this case, $z$ and $w$ take
values on the big oval of the spectral curve and
do change sign on $\partial B$. In this case, the vanishing of
the imaginary part of \eqref{circA} is equivalent to
the condition that the polygon inscribed in $B$ lifts to
a closed contour in $\R^3$, see the discussion after
the statement of Theorem \ref{Tins} below. The argument that 
was used to derive the real part of 
\eqref{circA} is no longer valid and, in fact, the 
real part of \eqref{circA} no longer holds. Instead, we have 
a different period condition that fixes the relative
height of the island with respect to other components
of $\partial\Omega$.

Let $\partial\Omega_1$ be the component of $\partial \Omega$ 
around which our island forms. Let $\partial\Omega_0$ be a 
different component of $\partial \Omega$ which respect to 
which the height of the island will be measured. Equation 
\eqref{arg} implies that 
$$
dh = - \frac{1}{\pi} \, \Im \alpha,
$$
and hence  
$$
\textup{height}\left(\partial\Omega_1\right) - 
\textup{height}\left(\partial\Omega_0\right) =
\frac1\pi \Im \int_{\partial\Omega_1}^{\partial\Omega_0} \alpha \,,
$$
where the integration is along a path connecting the two 
boundaries as in Figure \ref{isl_bub}. 

For contour as in 
Figure \ref{isl_bub}, at the endpoints we have $z=0$ and 
$w\in\R\setminus 0$. In particular, the forms $\log z\,  d\log w$ 
and $\log \zeta \, d\log \omega$ are integrable near the 
corresponding points of $P$ and $Q$. The same computation as
in proof of Proposition \ref{log_arias} shows that 
\begin{equation}
  \label{height_diff}
   \textup{height}\left(\partial\Omega_1\right) - 
\textup{height}\left(\partial\Omega_0\right) =
\frac1{c\pi} \Im \int_{\partial\Omega_0}^{\partial\Omega_1}
\log \zeta \, d\log \omega + \textup{const} \,. 
\end{equation}
The path of the integration in \eqref{height_diff} connects
the points of intersection of two different real ovals of 
$Q$ with the line $z=0$. We will call it the \emph{height} 
of one oval with respect to another.

In Section \ref{Qparam}, we will show that the logarithmic 
areas of compact ovals of $Q$, the heights of the 
noncompact ovals, and the points of intersection of 
$Q$ with the coordinate axes can be chosen as local 
coordinates on the space of nodal plane curves of given 
degree and genus. This suggests constructing the 
required curve $Q$ by deformation from the equally weighted,
 no islands case treated in 
the present paper. We hope to address this in a 
future paper.

\section{Proof of theorem \ref{t2}}

\subsection{Winding curves and cloud curves}

\subsubsection{Winding curves}

As we will see, the rational curve $Q$ in the
statement of Theorem \ref{t2} is not just any
rational plane curve, it has some very special
properties. We begin the proof of Theorem \ref{t2}
with a discussion of these special properties.

We say that a degree $d$ real algebraic curve $C\subset\RP$ is \emph{winding}  if
\begin{enumerate}
\item[(1)] it intersects every line $L\subset\RP$ in at least $d-2$
points counting multiplicity, and
\item[(2)] there exists a point $p_0\in \RP\setminus C$
called the \emph{center}, such that every line
through $p_0$ intersects $C$ in $d$ points.
\end{enumerate}

The center $p_0$ in the above definition is obviously not
unique: any point sufficiently close to $p_0$ can serve as a
center instead. The existence of a center implies that every winding
curve is hyperbolic, that is, it lies in the closure of the set
of curves having $[d/2]$ nested ovals encircling the
center. Much is known about hyperbolic curves, see
e.g.\ \cite{Spey,Vin}.

Winding curves are almost never smooth. They typically
have nodes, but may also have ordinary triple, quadruple etc., points as
well as tacnodes (points where two branches are tangent).
This, however, exhausts the list of possible singularities
in view of the following

\begin{Proposition}\label{sing_wind}
All singularities of a winding curve $C$ are real. Every branch of
$C$ through a singularity is real, smooth, and has contact of order
$\le 3$ with its tangent, that is, it has at most ordinary flexes.
The only double tangents of a winding curve
are ordinary tacnodes with two branches on the opposite
sides of the common tangent.
\end{Proposition}

\begin{proof}
Let $p$ be a complex singularity of $C$. The line joining $p$ with
the complex conjugate point $\bar p$ is real and meets $C$ in at least
$4$ complex points counting multiplicity---contradicting the definition.
Similarly, a complex
branch through a singularity
yields complex intersection points with any line that
just misses the singularity, in contradiction
with part (2) of the definition.

Now suppose the origin is
a real singularity of $C$ and
let a branch of $C$ be parameterized by
$$
x=t^a\,, \quad y=t^b+\dots \,,
$$
where $a<b$.
The line $y=0$ meets this branch with multiplicity $b$
whereas the nearby line $y=\epsilon$ meets it in $1$
real point if $b$ is odd and may miss it entirely if
$b$ is even. It follows that $b\le 3$.

The case $(a,b)=(2,3)$ of an ordinary cusp is ruled out
by considering a line from the cusp to the center $p$.
The order of contact with a nearby line through $p$
will drop by $2$.

It is easy to see that a line that is tangent to
two distinct points of $C$ can be moved so that it
loses $4$ real points of contact. The same is true
if two branches of $C$ are on
the same side of the common tangent, or tangent at a flex
(in which case the multiplicity of the intersection with the tangent
line is at least four greater than for that of certain nearby lines).
\end{proof}

\subsubsection{Moduli of rational winding curves}\label{mod_rat}

It follows from Proposition \ref{sing_wind} 
that a winding curve $Q$ is always 
\emph{immersed}, that is, the map 
\begin{equation}
  \label{normQ}
  f: \widetilde{Q} \to \bP^2
\end{equation}
from the normalization of $Q$ to the 
plane is an immersion. In this paper, we will focus on the winding curves 
of genus 
zero, in which case $\widetilde{Q}\cong\bP^1$.

Clearly, if $Q$ is irreducible, winding, and without tacnodes, then 
all small deformations of the map \eqref{normQ} remain winding. 
In other words, such curves form an open set of the real locus of the moduli 
space $\cM_0(\bP^2,d)$ of degree $d$ maps 
$$
f: \bP^1 \to \bP^2 
$$
considered up to a reparameterization of the domain. 
This moduli space is smooth of dimension $3d-1$, see \cite{FP}.  
This is because the projective plane $\bP^2$ is \emph{convex}, which, 
by definition, means that its tangent bundle is generated 
by global sections (concretely, the vector fields 
generating the $SL(2)$-action span the tangent space to $\bP^2$ at 
any point). 
The tangent space to $\cM_0(\bP^2,d)$
at the point $Q$ may be identified with the global sections 
$H^0(N_{\bP^2/Q})$, where $N_{\bP^2/Q}$ is the normal bundle (it is a bundle because
$Q$ is immersed). 

Suppose that the Newton polygon of $Q$ is nondegenerate, that is, 
suppose that $Q$ does not pass through the intersections of 
coordinate axes in $\bP^2$. Consider the monomials  in $Q(z,w)=0$
on the boundary of the Newton polygon. This gives $3d$ numbers
modulo overall scale, matching the dimension $3d-1$ of the moduli space. 
These outside monomials determine the $3d$ intersections of $Q$
with coordinate axes. In fact, these intersections should be 
properly considered as 
a point in $(S^d \, \C^*)^3$, subject to one constraint, which 
precisely identifies it with the outside coefficients of $Q(z,w)$.   

\begin{Proposition}\label{bdry_coord}
The outside coefficients of $Q(z,w)$ are local coordinates on $\cM_0(\bP^2,d)$
near an immersed curve $Q$ with a nondegenerate Newton polygon.
\end{Proposition}

\begin{proof}
 Let $\xi$ be a nonzero section of the normal bundle $N_{\bP^2/Q}$.
Let some branch of $Q$ intersect the axis $w=0$ at the point 
$(z_0,0)$ with multiplicity $m$. 
This intersection is preserved by $\xi$
if and only if $\xi$ vanishes at $(z_0,0)$ with the same order $m$. The 
differential 
\begin{equation}
  \label{def_eta}
     \eta = f^*(\iota_\xi \, d\log z \wedge d\log w)
\end{equation}
is a well-defined nonzero meromorphic differential on $\bP^1$. 
One checks that, in fact, it is everywhere regular. Indeed, 
near $(z_0,0)$ we have 
$$
w = a (z-z_0)^m + \dots \,, \quad \xi = b (z-z_0)^m \, \frac{d}{dw} + \dots
$$
where $a\ne 0$ and dots stand for $o((z-z_0)^m)$. Thus
$$
\eta  = - \frac{b}{az} \, dz + \dots 
$$
and we have produced a nonzero regular differential on a 
genus $0$ curve, a contradiction. 
\end{proof}

Three types of degeneration can happen to $Q$ in 
codimension one, see e.g.\ \cite{HM}. It can develop either:
\begin{enumerate}
\item[(i)] a cusp, or 
\item[(ii)] a tacnode, or 
\item[(iii)] an extra node, and, hence, become reducible. 
\end{enumerate}

All these degenerations mark the boundary of the winding 
locus in $\cM_0(\bP^2,d)$. By Proposition \ref{sing_wind}, 
a cusp can develop only when a loop of $Q$ ties tight
around the center $p$, as, for example, for 
$$
Q=z^2-(w-\epsilon)^2(w+\epsilon)\,, \quad  p=(0,0)\,, \quad 
\epsilon\to 0 \,.
$$
In this case, the frozen boundary, which is the dual curve 
of $Q$, escapes to infinity and so this case will not 
concern us in the present paper.

Curves with a tacnode or an extra node form 
codimension $1$ boundary strata of the winding locus. 
A real 
tacnode can deform to a pair of real nodes or a pair of
complex nodes; only in the first case does the curve 
remain winding. Similarly, only one of the two
smoothings of the new node is winding.

\subsubsection{Cloud curves}

The dual curve $C^\vee$ of a winding curve $C$
separates the dual projective plane into
the regions formed by those lines that intersect $C$ in $d$ and
$d-2$ points. We will call them the \emph{exterior} and
the \emph{interior} of $C^\vee$, respectively.
The line corresponding to the pencil of
lines though the center $p$ lies entirely in the exterior of $C^\vee$.
By making it
the line at infinity, the curve $C^\vee$ is placed into
the affine plane $\R^2$.

Since $C$ has no cusps,
$C^\vee$ has no inflection points and, hence, is locally convex except
for cusps. The cusps point into the interior of $C^\vee$, which
 makes the interior of $C^\vee$ resemble in shape a cloud,
see Figure \ref{cloudcurve}.
This is why we call the dual of a winding curve a \emph{cloud curve}.
Node that a cloud curve has no real nodes other than tacnodes,
for they would correspond to double tangents to $C$.

By construction, a cloud curve has a unique, up to
complex conjugation, complex (non-real)
tangent through any point in its interior.

\subsubsection{Higher genus}\label{Qparam}

Suppose that $Q$ is nodal curve of genus $g>0$ of the 
kind considered in Section \ref{is_b}, not 
necessarily winding. The space of nodal plane curves of 
given degree $d$ and genus $g$ is well known to be 
smooth of dimension $3d-1+g$, 
see \cite{HM}. 
As in Section \ref{mod_rat} above, the intersections 
with coordinate axes give $3d-1$ degrees of freedom. 
The remaining $g$ dimensions come precisely from the $g$-dimensional 
space of holomorphic differentials as in the proof 
of Proposition \ref{bdry_coord}. 

\begin{Proposition}
The points of intersection with coordinate axes, the 
logarithmic areas of compact ovals, and the heights of 
the noncompact ovals, can be chosen as local coordinates at $Q$ 
on the moduli space of genus $g$ degree $d$ nodal plane curves. 
\end{Proposition}

\begin{proof}
Consider the tangent space at $Q$ to the space of 
all genus $g$ curves with the same outside coefficients as $Q$. 
The proof of Proposition \ref{bdry_coord} identifies this 
tangent space with the 
space of regular differentials $\eta$ on the curve $Q$. 
Moreover, by \eqref{def_eta}, the derivative
 of the logarithmic area 
(resp.\ height) of an oval $O_i$ of $Q$ in the direction of $\eta$
is given by the real (resp.\ imaginary) part of $\int \eta$ 
along the corresponding path $\beta_i$ in Figure \ref{isl_bub}.

The contours corresponding to the compact ovals of $Q$
are closed and invariant under complex conjugation. Thus the 
corresponding periods are automatically real. As to the 
contours corresponding to the noncompact ovals, we can 
make them closed by taking their union with the complex
conjugate contour (the dashed line in Figure \ref{isl_bub}). 
This makes the contour anti-invariant with respect to 
complex conjugation and thus picks out precisely the 
imaginary part of the original integral. Finally, since 
the contours in Figure \ref{isl_bub} are disjoint, the 
period matrix $\int_{\beta_i} \eta_j$ is nondegenerate. 
\end{proof}

\subsection{Inscribing cloud curves in polygons}

Let $\Omega$ is a polygon
in the plane, not necessarily convex, formed by $3d$ segments with slopes $0,1,\infty$,
cyclically repeated as we follow the boundary in the
counterclockwise direction. Here $d=2,3,\dots$. For any constant $c$, the image
of $\Omega$ under the map
$$
\Ex: (x,y) \to  (e^{cx},e^{cy})
$$
is a polygon in first quadrant $\R_{\ge 0}^2$. The sides of $\Ex(\Omega)$
are formed by lines passing through the vertices of
$\R_{\ge 0}^2 \subset \RP$.

We will say that a real algebraic
curve $R$ is \emph{inscribed} in the polygon $\Ex(\Omega)$
if $R$ is a simple closed curve contained in $\Omega$ and
tangent to lines forming the sides of $\Omega$ in their
natural order. Note that near a nonconvex corner,
an inscribed curve may be tangent to the line
forming a side of $\Omega$ but not to the side itself.

Recall that, by definition, the class of a curve is the
degree of the dual curve. Our main result in this section is the following

\begin{Theorem}\label{Tins} Let $\Omega$ be a polygon as above. A class $d$
rational cloud curve can be inscribed in $\Ex(\Omega)$ if and only if
$\Omega$ is feasible. If it exists, the inscribed curve $R$ is
unique.
\end{Theorem}

The tangent vector to $R$ rotates $d$ times as we go once around
the curve. It follows that $R$ has $d-2$ real
cusps, or one cusp per each non-convex corner of $\Omega$. Pl\"ucker formulas,
see e.g.\ page 280 of \cite{GH}, imply that
the degree of $R$ equals $2d-2$. In absence of tacnodes, in addition to the $d-2$
real cusps, the curve $R$ has $2d-4$ complex cusps and
$2(d-2)(d-3)$ complex nodes.

The curve $R$ being tangent to $3d$ lines imposes $3d$ incidence
conditions on the degree $d$ dual curve $Q=R^\vee$. As discussed 
earlier, one of these
conditions is redundant. 
In fact, this redundancy translates geometrically
into the condition that $\Omega$
is the projection of a \emph{closed} contour in $\R^3$.
To see this, note that the condition of being a closed contour
is equivalent to the condition that, as the boundary of $\Omega$ is
traversed, the total signed displacement along each of the three
edge directions is zero. If the vertical edges are
at coordinates $x=x_1,\dots,x_d$, the horizontal edges are at
$y=y_1,y_2,\dots,y_d$ and the slope-$1$ edges are at
$x-y=z_1,z_2,\dots,z_d$, then the displacements along
horizontal edges are $y_1+z_1-x_1,y_2+z_2-x_2,\dots,y_d+z_d-x_d$.
To be closed the sum of these displacements must be zero.
The fact that the displacements in the other two directions are zero gives an
identical relation.
The equation 
$$
\sum_{i=1}^d y_i+z_i-x_i=0
$$ 
is the
logarithm of the condition that the intersection points of a plane curve
with the three coordinate axes has product $1$.

Also note that while imposing $3d-1$ incidence conditions
on a rational plane curve of degree $d$ gives generically
a finite set of possibilities, the number of possibilities
grows very rapidly. For example, there are 26312976
rational curves of degree 6 through 17 general points in
the plane (all of these numbers were first determined
by Kontsevich, see e.g.\ \cite{FP}). This shows that solving
the incidence equation requires some care. In fact,
our existence proof can be turned into a practical
numerical homotopy procedure for finding the unique
inscribed curve.

\subsection{Proof of Theorem \ref{Tins}}\label{sTins}

\subsubsection{Strategy}
Our strategy will be deformation to $c\to\pm\infty$ limit, in
which tropical algebraic geometry takes over. This is essentially
the patchworking method of O.~Viro, see e.g.\ \cite{ItVir}. A closely
related approach was employed, for example, in \cite{Mikh2} to
construct curves of given degree and genus passing through given
points in the plane.

\begin{figure}[htbp]
 \begin{center}
  \rotatebox{180}{\scalebox{0.5}{\includegraphics{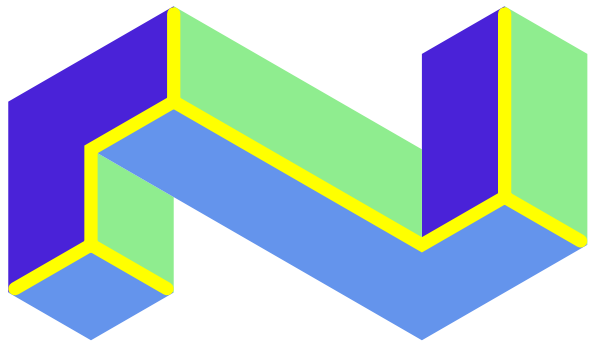}}}
  \end{center}
 \caption{Surface enclosing the minimal volume
\label{fig_max_surf}}
\end{figure}

{F}rom the point of view of random surfaces, letting
$c\to\pm\infty$ means imposing an extreme volume constraint.
As $c\to+\infty$, the limit shape becomes the unique piecewise linear
surface that minimizes the volume enclosed, see
an example in Figure \ref{fig_max_surf}.

In the $c\to\pm\infty$ limit, the inscribed curve collapses
to (a part of) the corner locus (locus where the surface is not smooth)
of this
piecewise linear surface, an example of which is
plotted in yellow in Figure \ref{fig_max_surf}.
The relevant parts of the corner locus
are convex edges for volume maximizers and  concave
edges for minimizers (as seen from above the graph).

Our goal is to reverse this degeneration, that is, starting with the
concave corner locus of the volume minimizing surface, we construct
a certain tropical curve $Q_T$, the would-be dual of the inscribed curve.
Then we show that $Q_T$ can be deformed to a rational degree $d$
winding curve $Q(x,y;c)=0$ whose dual curve
$R(x,y;c)=0$ is inscribed in $\Ex(\Omega)$ for all sufficiently
large $c$. Finally, we prove that these curves can be further
deformed to any finite value of $c$.

\begin{figure}[!htbp]
  \centering
  \rotatebox{180}{\scalebox{0.64}{\includegraphics{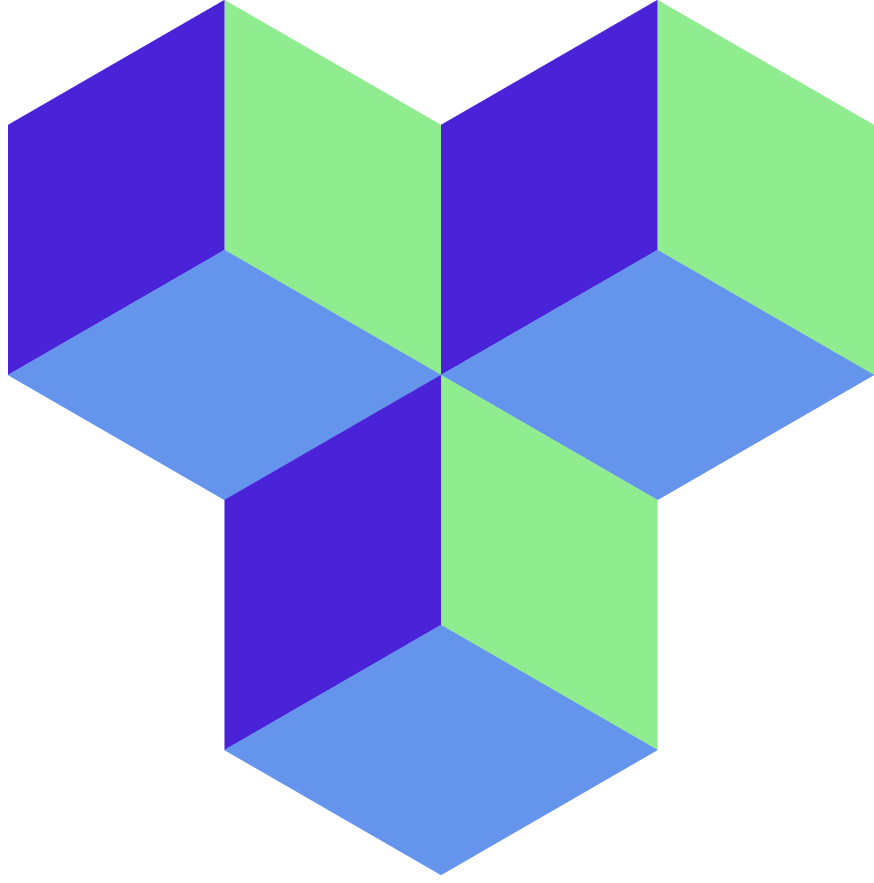}}}
  \rotatebox{180}{\scalebox{0.64}{\includegraphics{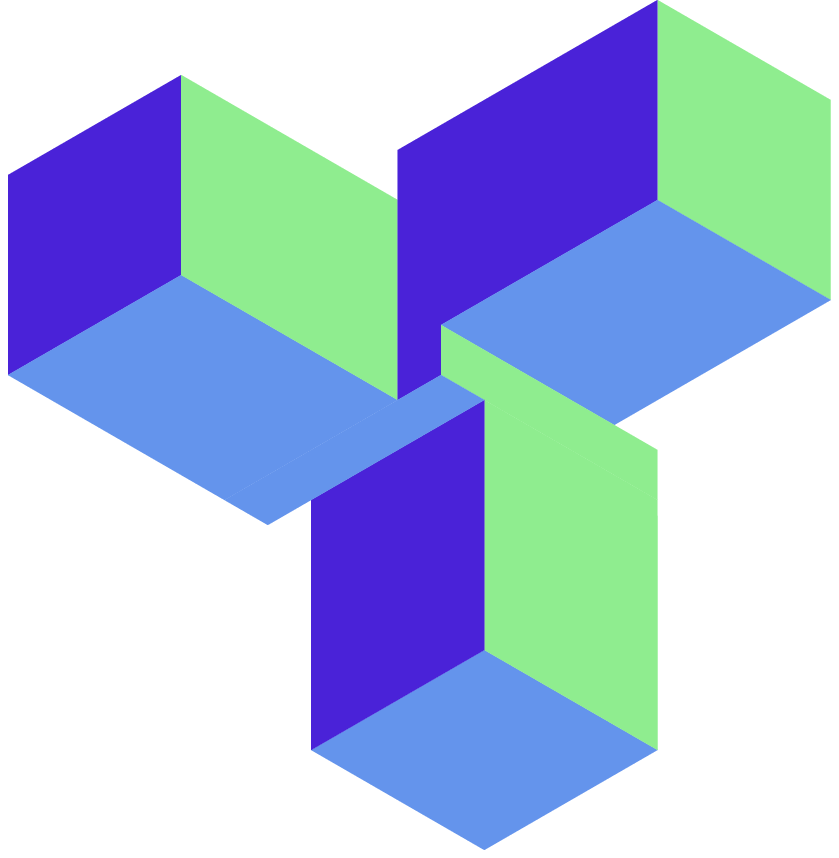}}}
  \caption{A nongeneric polygon $\Omega$ and a nearby generic one.
  \label{fig_gen}}
\end{figure}

\begin{figure}[!htbp]
  \centering
  \rotatebox{180}{\scalebox{0.64}{\includegraphics{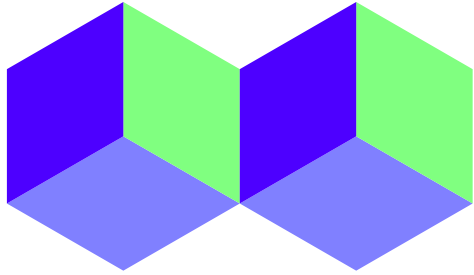}}}
  \caption{A polygon with a taut path (the vertical bisector).
  In a tiling, no tile crosses the taut path.
  \label{tautpath}}
\end{figure}

In order for the very first step to work, we need to assume that
$\Omega$ is feasible, otherwise there is no volume maximizing
surface to begin with. We further assume that $\Omega$ is
generic, that is, there are no accidental relations
between its side lengths. In particular,
for generic $\Omega$, the triple points of the concave corner locus
of the volume minimizers are
local minima, see the illustration in Figure \ref{fig_gen}.
Moreover, genericity implies that the maximizing surface
lies strictly above the minimizing surface, except at the boundary.
This implies that there are no ``taut'' paths, that is, paths in the interior which
are present in every surface. See Figure \ref{tautpath} for an example
of a taut path.

\subsubsection{The tropical curve}

\begin{figure}[!htbp]
  \centering
  \scalebox{0.4}{\includegraphics{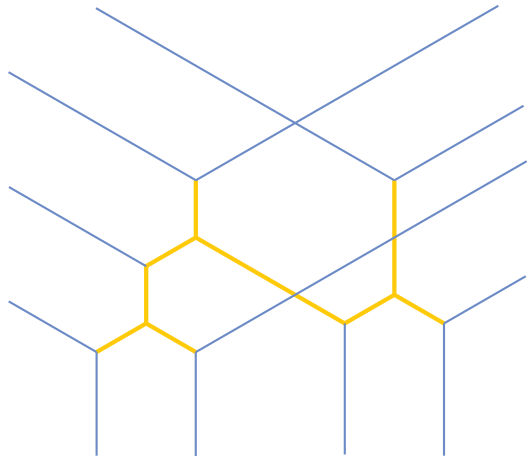}}
  \caption{The tropical curve associated to the surface in Figure \ref{fig_max_surf}.
  \label{fig_tropic}}
\end{figure}

The construction of the tropical curve is very simple:
take the concave corner locus of the volume minimizer and
extend univalent and 2-valent
vertices to 3-valent ones as in Figure \ref{fig_tropic}, where
the newly added rays are shown in blue (thinner lines).

The tropical curve $Q_T$ thus obtained is a genus zero degree $d$
tropical hyperbolic  curve, see \cite{Spey} where properties of
such curves are discussed in detail.

Consider the connected components of the complement of $Q_T$,
which we will call \emph{chambers}. There are $\binom{d+2}{2}$
chambers and they naturally correspond to lattice points in
the triangle with vertices $(0,0)$, $(d,0)$ and $(0,d)$.
The meaning of this triangle
for us is that it is the Newton polygon of the degree $d$
curve $Q(x,y;c)$.

The dual of the graph $Q_T$
is, combinatorially, a tessellation of this triangle
with unit size triangles and rhombi (recall that we assume $\Omega$ is generic).
Each rhombus in the tessellation will correspond
to a node of $Q(x,y;c)=0$; there will be $\binom{d-1}{2}$ of such.

Note that triple points of $Q_T$ come in two different orientations,
and triple points of one of the two orientations correspond 
bijectively to local maxima of the concave
corner locus.

\subsubsection{Construction of $Q$}
We will first construct an approximation
\begin{equation}
  \label{defQ0}
   Q^0(x,y;c) = \sum_{i+j\le d} q^0_{ij} \, x^i y^j \,,
\end{equation}
to the true winding curve $Q$.

Recall that the monomials of $Q$ correspond to chambers of $Q_T$.
For example, monomials that differ by a factor of $x$
correspond to adjacent chambers separated by a vertical line
(which sometimes can have zero length).
Let $x=\alpha_{ij}$ be the line separating the chambers
corresponding to the monomials $x^i y^j$ and $x^{i+1} y^j$.
Then we require that
$$
\frac{q^0_{i+1,j}}{q^0_{i,j}} = e^{-c\alpha_{ij}} \,.
$$
Similarly, for monomials that differ by factor of $y$,
we set
$$
\frac{q^0_{i,j+1}}{q^0_{i,j}} = e^{-c\beta_{ij}}
$$
if the line
separating the chambers is $y=\beta_{ij}$.
For monomials which share a diagonal edge $x-y=\gamma_{ij}$ we have
$$
\frac{q^0_{i,j+1}}{q^0_{i+1,j}} = e^{-c\gamma_{ij}} \,.
$$
Since at a trivalent vertex we have
$$\beta_{ij}-\alpha_{ij}=\gamma_{ij},$$
this rule defines the curve $Q^0=0$ uniquely and unambiguously
up to a common factor.
The irrelevant common factor may be chosen so that all coefficients
of $Q^0$ are real and positive.

By construction, as $c\to\infty$ the amoeba of the curve $Q^0$, scaled
by $c$,
converges to the tropical curve $Q_T$. In particular, $Q^0$
intersects the coordinate axes approximately at the points
\begin{equation}
  \label{inf_points}
  (1,0,-e^{c\alpha_i}), \quad (0,1,-e^{c\beta_i}), \quad (1,-e^{-c\gamma_i},0) \,,
\end{equation}
where
$$
x = \alpha_i, \quad y=\beta_i, \quad x-y=\gamma_i, \quad i=1,\dots,d\,.
$$
are the lines forming the $3d$ sides of $\Omega$
By a small $c$-dependent change of the original curve $Q_T$, we can make
the curve $Q^0$ defined by \eqref{defQ0}
pass precisely through the points \eqref{inf_points}.

A  small deformation of a quadruple point
of $Q_T$ (into a pair of triple points connected in one of two ways)
leads to the change of the topology
of the real locus of $Q^0$ for all large $c$. There is a
$c$-dependent deformation of $Q_T$ for which every
quadruple point becomes a real node, see \cite{Mikh2}. Thus, for all $c\gg 0$,
 we have constructed a genus zero curve passing through the points
\eqref{inf_points}. This is our curve $Q(x,y;c)$.

\subsubsection{Intersections with lines}
\begin{figure}[!htbp]
  \centering
  \scalebox{0.64}{\includegraphics{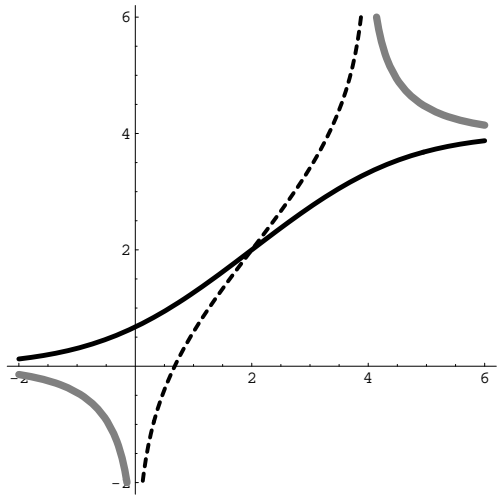}}
  \caption{The curves $Q(\pm e^{x},\pm e^{y})=0$ for
$Q(z,w)=1+z+w+e^{-4}zw$.}
  \label{fig_twist}
\end{figure}

\begin{figure}[!htbp]
  \centering
  \scalebox{0.4}{\includegraphics{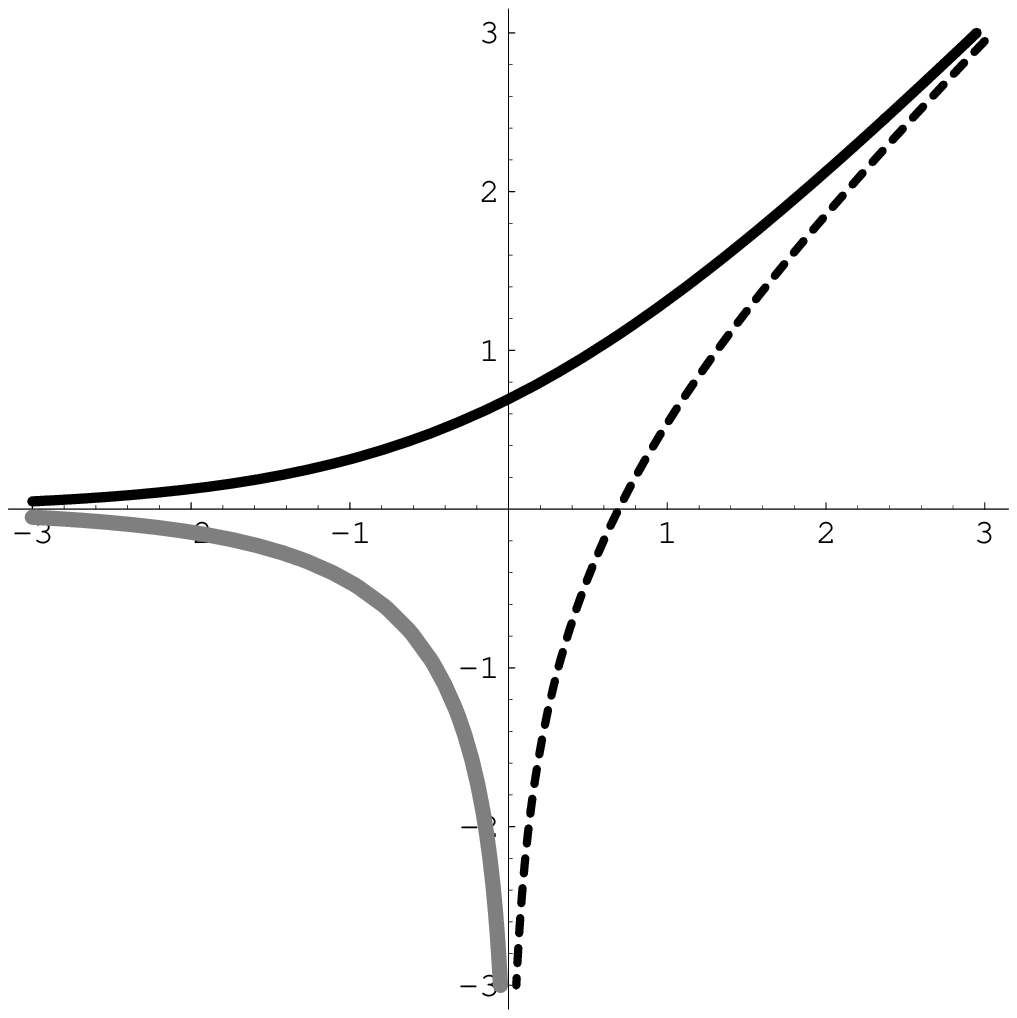}} \quad \scalebox{0.4}{\includegraphics{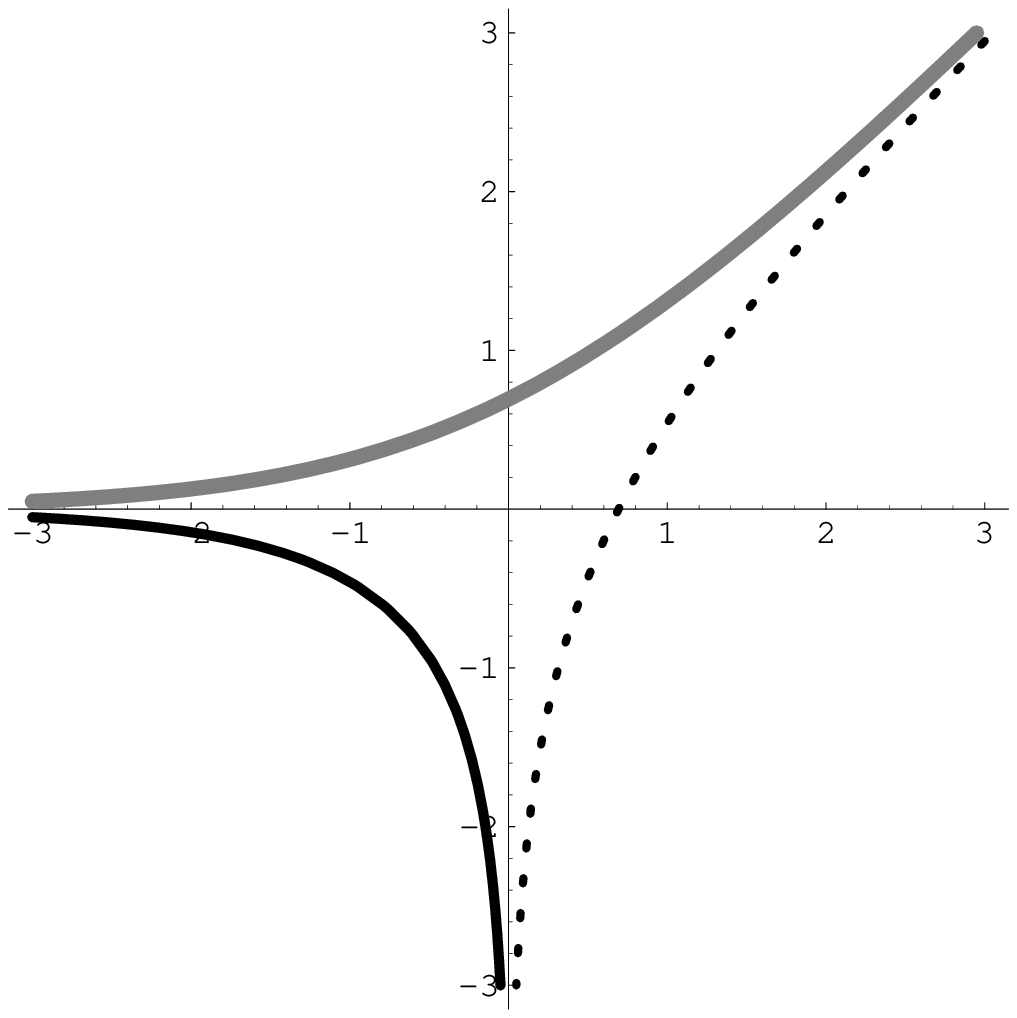}}
  \caption{The curves $Q(\pm e^{x},\pm e^{y})=0$ for
$Q(z,w)=1+z+w$ and $Q(z,w)=1-z+w$.}
  \label{fig_lines}
\end{figure}

\begin{figure}[!htbp]
  \centering
  \scalebox{0.6}{\includegraphics{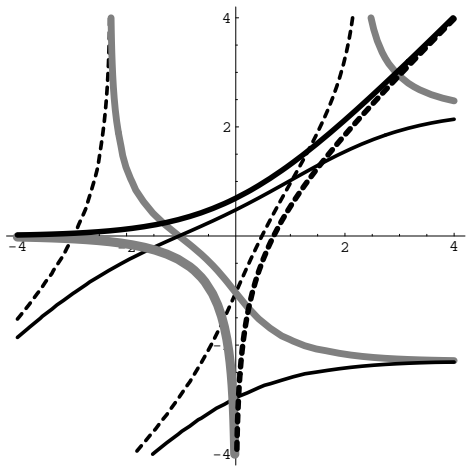}}
  \caption{How to lose four intersections. The $Q$ curve here
  is in thinner colors, the line in thicker colors.
  \label{lose4ints}}
\end{figure}

Our next step is to prove that the curve $Q(x,y;c)$ is winding.
This involves understanding its real locus. We know that
the curves
\begin{equation}
  \label{Qpm}
  Q(\pm e^{cx}, \pm e^{cy}; c) =0
\end{equation}
approach $Q_T$ as $c\to\infty$. One of these four curves, the
one that corresponds to the $(+,+)$ choice of signs is empty
because $Q$ has only positive coefficients. The shape of the
other three curves can be inferred from simplest example
of a conic, shown in Figure \ref{fig_twist}. The
colors in Figures \ref{fig_twist}, \ref{fig_lines} and \ref{lose4ints} are explained in the
following table:
\begin{center}
\begin{tabular}{c|cccc}
quadrant  & $(+,+)$ & $(+,-)$ & $(-,+)$ & $(-,-)$ \\
\hline
color & dotted & black & dashed & gray
\end{tabular}
\end{center}
The essential feature seen in Figure \ref{fig_twist} is that
the black and dashed curves cross over near the middle of
the corresponding edge of the tropical curve. (Note that
for clarity in some of the figures we use rectilinear
coordinates and in others, coordinates with the $3$-fold symmetry. 
Hopefully these coordinate changes will not confuse the reader).

\begin{figure}[!htbp]
  \centering
  \scalebox{0.5}{\includegraphics{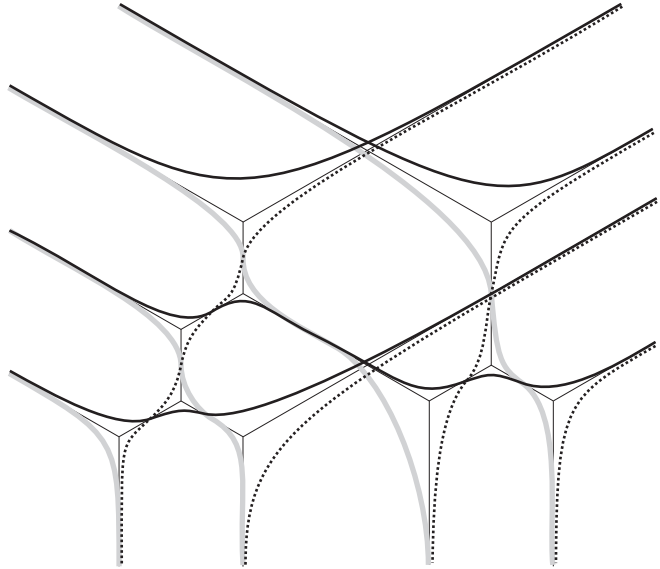}}
  \caption{This is how the curves $Q(\pm e^{cx}, \pm e^{cy}; c)=0$
look like for $c\gg 0$
for the tropical curve in Figure \ref{fig_tropic}.}
  \label{fig_real_locus}
\end{figure}

This same feature holds in general since for $c$ sufficiently large, the structure
of the curves near a compact edge of $Q_T$
only depends on the coefficients of $Q$ corresponding to the
four chambers around that edge: the other coefficients are exponentially
small compared to these.
It follows that the curves \eqref{Qpm}, in our running
example, look like the curves in Figure \ref{fig_real_locus}.

Now we turn to intersecting the curves $Q(x,y;c)$ with lines.
There are two kinds of lines, those that intersect the positive
quadrant and those that don't. The latter ones can be defined
by an equation with positive coefficients. The difference
between the two kinds of lines is illustrated in Figure \ref{fig_lines}.
As we will see only lines that miss the positive quadrant have
a chance to intersect $Q(x,y;c)$ in fewer than $d$ points.

A basic property of any generic tropical curve, and of $Q_T$ in
particular, is that it intersects a generic
tropical line in $d$ points, where $d$ is the degree. Generic
here means that the triple point of the line is not a point of
the tropical curve. It is easy to see that in this generic
situation there are still $d$ points of intersection of the
nearby nontropical curve $Q$ with a nearby line: transverse intersections
are preserved under deformations.

If a line is not generic, some of these points of intersection
may be lost. Note that real points have to be lost in pairs,
so, for example, if the triple point of the line lands on a
(noncompact) ray of $Q_T$, there are still $d$ real
points of intersection for nearby curves (this follows because
the point at infinity has two colors for both $Q$ and the line
and so this point is still a point of intersection). However,
if the triple point of the line
lands on a compact segment of $Q_T$, two points of intersection
may disappear. If the triple point of the line lands
on a triple point of $Q_T$ of the opposite orientation, two points of intersection
(at most) may disappear.
Finally, if the triple point of the line is a triple
point of $Q_T$, with the same orientation, and all incident
segments of $Q_T$ are compact, then as many as
$4$ real points of intersection may be lost (Figure \ref{lose4ints}).

A simple but crucial observation is that, by construction, the
curve $Q_T$ has no triple points of the same orientation as
the tropical line and incident to 3 compact edges. This is
because, as discussed earlier and illustrated in Figure \ref{fig_gen},
triple points of compact edges correspond to local minima of
the volume minimizer for generic $\Omega$.

Furthermore, it is elementary to check that if the line
intersects the positive quadrant then it has $d$ points
of intersection with $Q$. Similarly, any tropical line
with a triple point sufficiently far away from the amoeba of $Q_T$
meets $Q$ in
$d$ points. Thus the curve $Q$ is winding.

The dual cloud curve $R=Q^\vee$, which, recall, is convex and 
non-sef-intersecting, is approximately the
yellow convex corner locus from Figure \ref{fig_max_surf}.
This is because the tropical lines whose triple points lie on
a compact edge of $Q_T$ are exactly those which
lead to fewer than $d$ points of
intersection. The curve $R$ is inscribed in $\Omega$ because, if it were
protruding it would have more, say, vertical tangents than
we constructed. This is impossible since more than $d$ vertical
tangents means more than $d$ points of intersection of $Q$ with
a certain line.

\subsubsection{Deforming the winding curve}\label{cloc}

We will now show that the inscribed curve that we constructed
for $c\gg 0$ can be deformed to any given value of $c$.
Changing $c$ produces a particular $1$-parameter family 
of polygons into which a cloud curve is to be inscribed. 
It is more convenient, in fact, to consider the general
variation of the polygon $\Ex(\Omega)$. 

In the set of all polygons $\Omega$ with $3d$ disjoint 
edges (that is, no part of $\partial \Omega$ is traversed twice)
consider the set of all feasible ones. Denote the interior
of this feasible set by $\Upsilon$. Each connected component of 
$\Upsilon$ is naturally an open set of $\R^{3d-1}$. 
Let $\Upsilon_0 \subset \Upsilon$ be the set of those 
$\Omega$ for which there exists an irreducible cloud 
curve $R$ without tacnodes inscribed in $\Ex(\Omega)$ with 
$c=1$. Our goal is to prove that $\Upsilon_0 = \Upsilon$. 

We know that the intersection of $\Upsilon_0$ with each 
connected component of $\Upsilon$ is nonempty. In fact, 
if $\Omega$ is generic, then $c\Omega$, $c\gg 0$,  
is contained in $\Upsilon_0$. 

Observe that $\Upsilon_0$ is open. 
This follows from Proposition \ref{bdry_coord} because
changing $\Omega$ amounts to moving the points of 
intersection of $Q$ with the coordinate axes. 

We claim that $\Upsilon_0$ is also closed. Indeed, let 
$\Omega$ be a limit point of $\Upsilon_0$. Since the 
set of winding curves is closed and $R$ remains 
bounded, there will be an 
inscribed cloud curve $R$ in $\Ex(\Omega)$, but it might 
be reducible or tacnodal, or both. Since the map from $R$ 
to the circumscribed polygon is continuous, it is 
enough to show that the codimension $1$ degenerations
cannot happen, that is, it is impossible to 
acquire exactly one tacnode or one extra node. 

\begin{figure}[!htbp]
 \begin{center}
  \scalebox{0.45}{\includegraphics{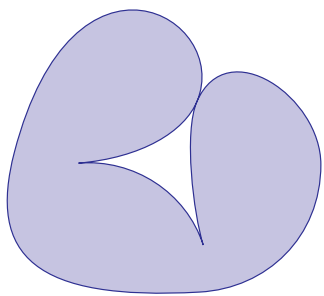}}
  \end{center}
 \caption{A cloud curve develops a tacnode} 
\label{swall}
\end{figure}

If the curve $R$ develops a tacnode this means that 
a part of the frozen region is swallowed by the liquid
region as in Figure \ref{swall}. This is impossible because,
for example, there must be a point of tangency 
with the boundary between any two cusps of $R$. 

\begin{figure}[!htbp]
  \centering
  \scalebox{0.64}{\includegraphics{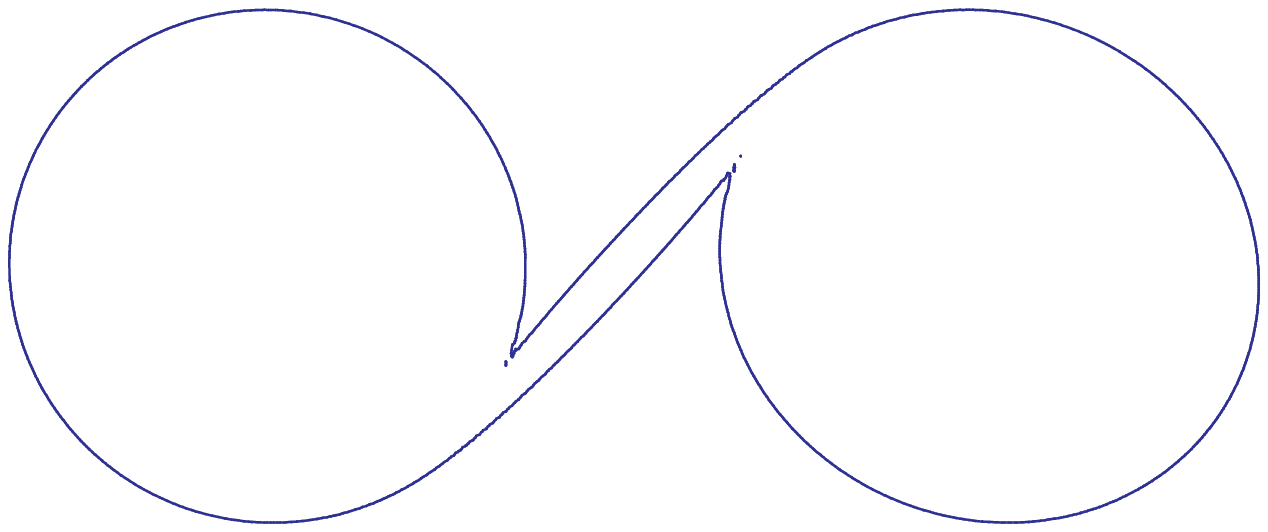}} \quad
   \scalebox{0.3}{\includegraphics{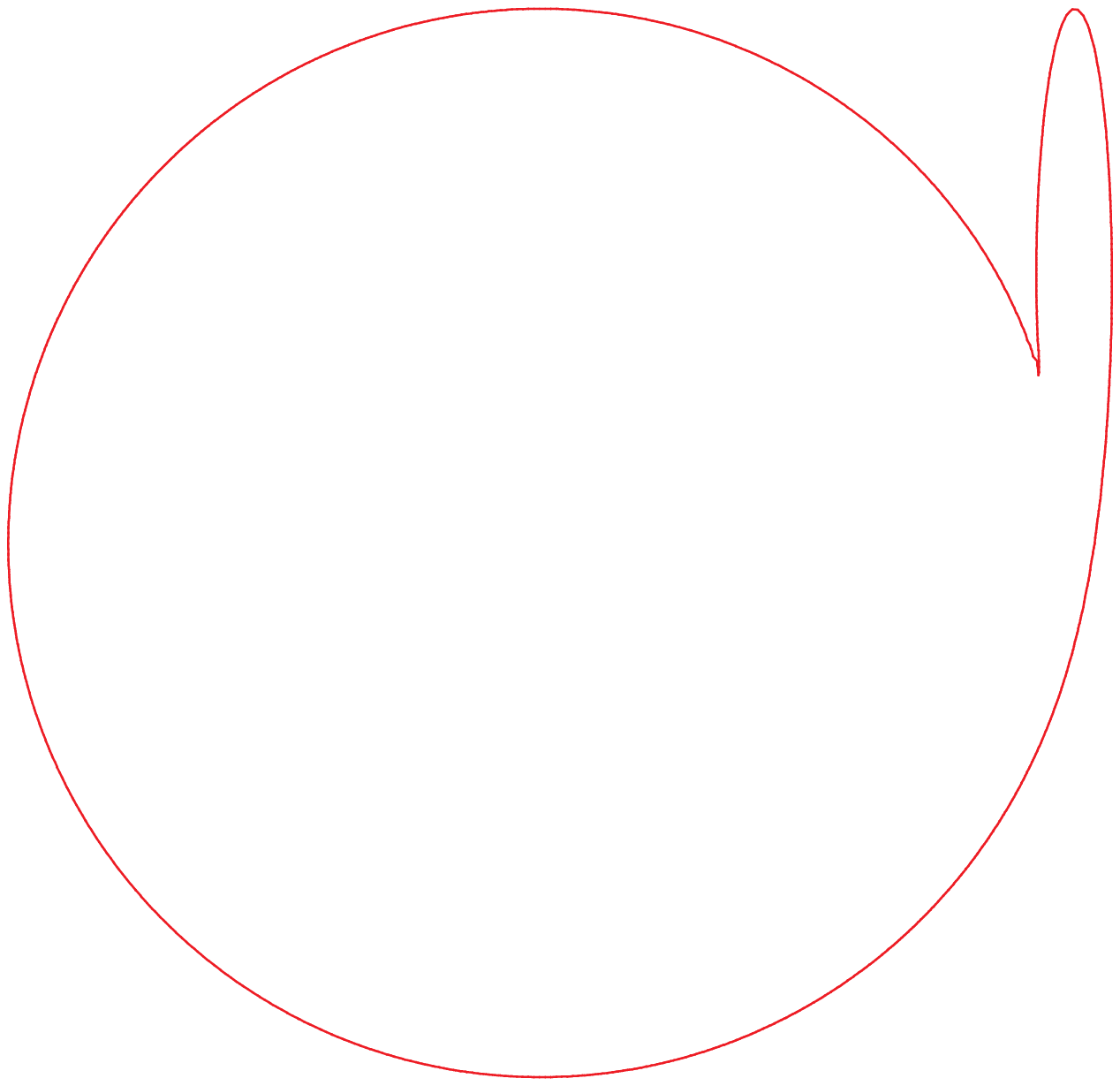}}
  \caption{\label{f_degen}A cloud curve becomes reducible}
\end{figure}

The other potential degeneration 
is when $Q$ develops an extra node
(and, hence, becomes reducible).
First suppose that neither component of the
degenerate curve is a line. In this case,
the dual curve $R$ becomes the
union of two cloud curves and a double line
tangent to both of them. Nearby cloud curves
look like a union of the two disjoint
cloud curves connected by a thin elliptical piece along a double
tangent, see illustration in Figure \ref{f_degen} on
the left. This, for
example, is what happens if we squeeze the polygon
in Figure \ref{fig_max_surf} in the middle until the middle portion goes to zero
width. This example, however, is \emph{not} generic because
the linear pieces of the limit shape  on two sides
of the collapsing ellipse have different slopes and meet along
a line in a coordinate direction. Generically,
the double tangent is not a line in a coordinate
direction and these linear pieces (which then have the same slopes)
merge together.
This leads to a taut path in the polygon and, hence, cannot happen
in our situation.

The other possibility is when $Q$ degenerates to
a union of a winding curve $Q_1$ and a line $Q_2$.
The line $Q_2$ corresponds
to a point $p_2$ of the dual plane and the collapsing
ellipse follows the tangent from $p_2$ to $Q^\vee_1$,
this is illustrated in Figure \ref{f_degen} on the right.
This degeneration can only happen if an edge of $\Omega$
has length tending to zero.

Note that in codimension two one can have a degeneration
in which two cusps of the frozen boundary come together
and merge in a tacnode, reminiscent of the Henry Moore's
\emph{Oval with Points}. This corresponds in Figure \ref{f_degen},
left part, to the thin ellipse having zero size.

\begin{figure}[htbp]
 \begin{center}
 \scalebox{0.4}{\includegraphics{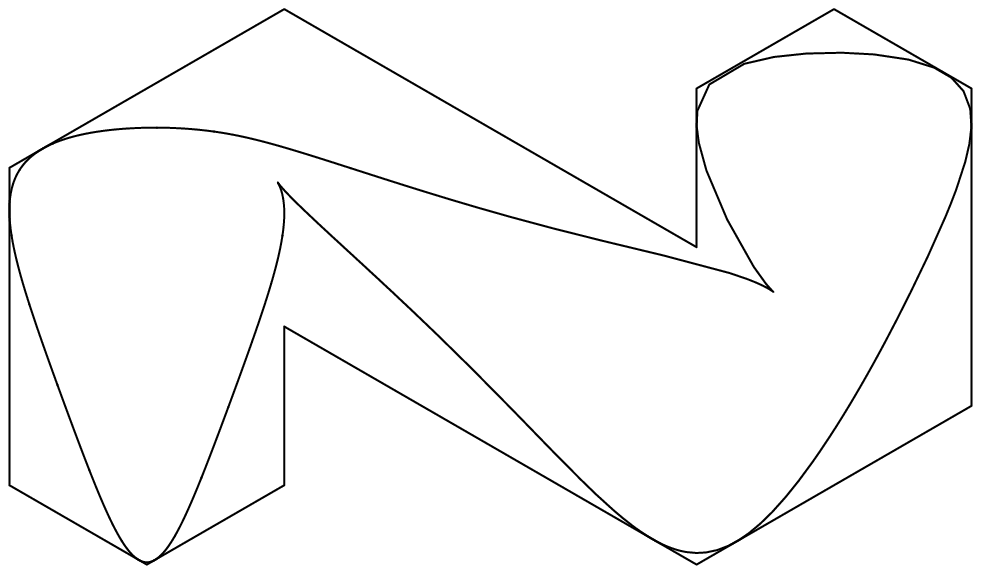}\hskip.3in
 \includegraphics{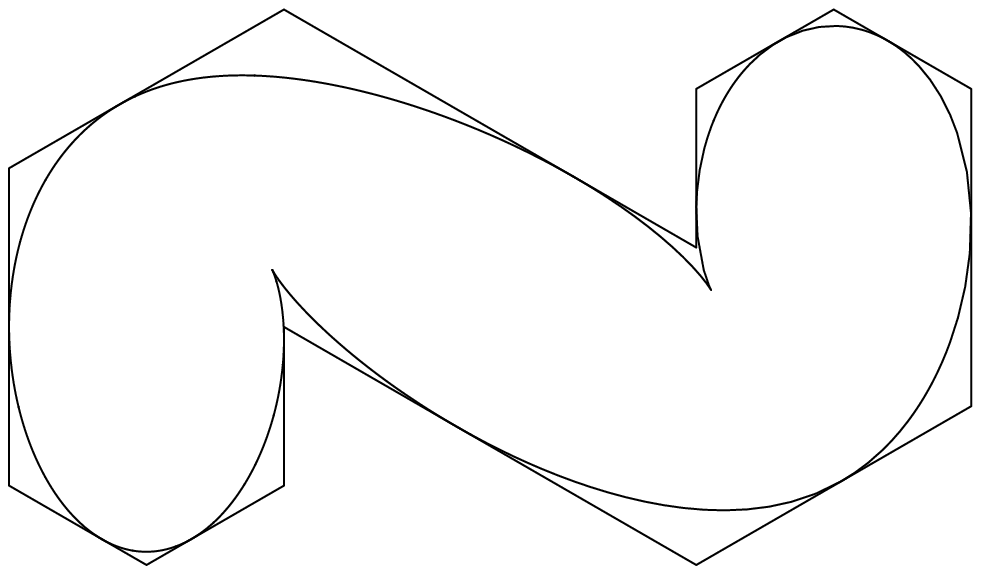}\hskip.3in\includegraphics{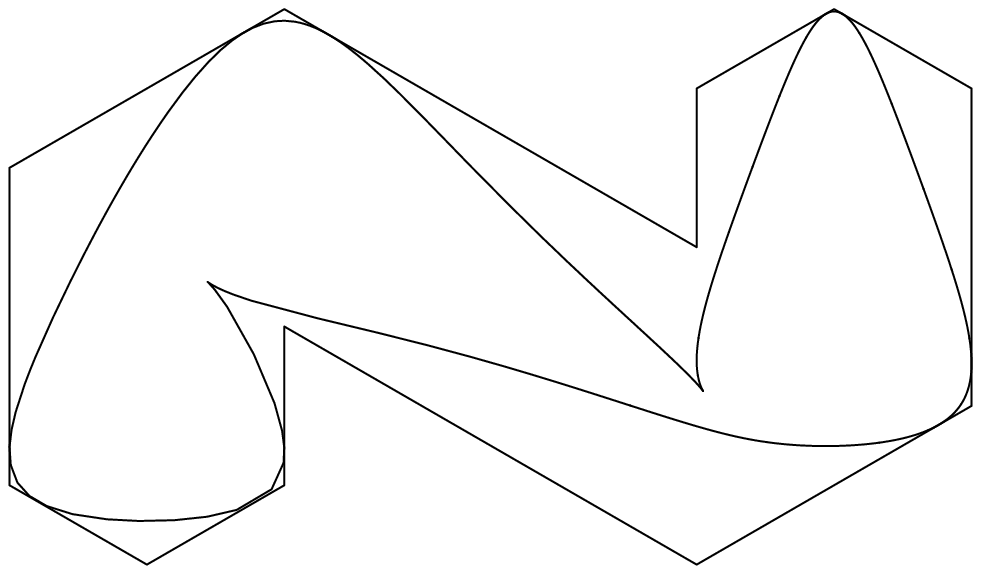}}
 \end{center}
\caption{Frozen boundaries (at $c<0, c=0$ and $c>0$) for the polygonal region in Figure \protect\ref{fig_max_surf}.}
\label{cloudcurve}
\end{figure}

Finally, there is practical side to the above
deformation argument. Namely, it allows us to find 
the inscribed curve by \emph{numeric homotopy}, that is, 
by solving the equations using Newton's algorithm with 
the solution of the nearby problem as the starting 
point. An 
example of a practical implementation of this can be 
see in Figure \ref{cloudcurve}. 

\subsubsection{Uniqueness}
It remains to prove that the inscribed curve is
unique and that the polygon is feasible if the
inscribed curve exists. Given the inscribed
curve $R$, let $\hs$ denote the height
function obtained from it by the procedure
of Section \ref{s_alg}. This is a well-defined
function with gradient in the triangle \eqref{triang}.
Therefore $\Omega$ is feasible. The inscribed
curve is unique because by the results of
Sections \ref{cloc} it has
a unique deformation to the unique tropical inscribed
curve.

\section{Global minimality}\label{sgmin}

As before, let $\hs$ denote the height function
constructed from the inscribed cloud curve $R$.

\begin{Theorem}
The height function $\hs$ is the unique minimum for
the variational problem \eqref{varp}.
\end{Theorem}

\begin{proof}
By strict convexity of the surface tension, there is
a unique minimizer.
Therefore we need to prove that
\begin{equation}
  \label{gh}
   \cE(g) \ge \cE(\hs)
\end{equation}
for any function $g$ satisfying the boundary conditions and
with gradient in \eqref{triang}. Since the integrand in \eqref{varp}
is bounded, we can freely modify it on sets of arbitrarily
small measure. In particular, it is enough to prove \eqref{gh}
for functions $g$ such that $g=\hs$ in a neighborhood of
the vertices of $\Omega$.

For any such $g$, we will construct a sequence  $\{h_n\}$ converging to
$h$ such that
\begin{itemize}
\item[(i)] $\cE(h_n) \to \cE(\hs)$, as $n\to\infty$\,,
\item[(ii)] $\cE$ has a directional derivative at $h_n$ in the direction of $g-h_n$\,,
\item[(iii)]  this directional derivative is nonnegative for all $n\gg 0$ \,.
\end{itemize}
By convexity of $\cE$, the last condition implies that $\cE(g) \ge \cE(h_n)$
for all $n\gg 0$, from which \eqref{gh} follows.

Now we proceed with the construction of the sequence $\{h_n\}$. In the
liquid region we will take $h_n=\hs$, except in a very thin strip along
the frozen boundary. Inside the facet, $h_n$ will be a smooth function very
slightly deviating from $\hs$ and equal to it in a fixed neighborhood of
the vertices of $\Omega$. Note that
the facets of $\hs$ are of two types: the ``bottom'' facets, for which
$g\ge \hs$ for any allowed function $g$, and the ``top'' facets, where
the inequality is reversed. For example, the bottom facet in Figure
\ref{fcard} is also a bottom facet according to the above definition.
For concreteness, we will give a construction of $h_n$ on a bottom
facet; the case of a top facet is the same, with obvious modifications.

The construction of Section \ref{sTins} give a family of solutions of
the Euler-Lagrange equation \eqref{EL} parameterized by a real
constant $c$. Since we will need all of them for the argument
that follows, we will use $c_0$ to
denote the particular value of the Lagrange multiplier in \eqref{varp}
for which we wish to prove that $\hs$ is a minimizer. Note that 
the height at any point is a decreasing function of $c$. Indeed, the 
$c$-derivative $\phi = \frac{\partial \hs}{\partial c}$ 
of the height function solves the linearization of 
\eqref{EL}, which has the form
$$
\dv G \cdot \nabla \phi = 1 \,, 
$$
with a positive definite matrix $G=\nabla^2\sigma(\nabla \hs)$ in the 
liquid region and zero boundary conditions on the frozen boundary. 
The condition $G\ge 0$ guarantees $\phi$ has no local maxima and 
hence $\phi\le 0$. 

Consider a horizontal bottom facet of $\hs$. Its boundary is formed by a part
of $\partial\Omega$ and a part of the frozen boundary. Let us
denote that part of the frozen boundary $F_{c_0}$ and call it
the \emph{frozen front}. For $c$
below $c_0$, the monotonicity $\frac{\partial \hs}{\partial c}\le 0$ 
implies the frozen front $F_c$ moves inside the facet. It 
sweeps the entire facet as $c\to -\infty$.

By our assumption on $g$, we can find $C\ll 0$ such that $g=\hs$
outside $F_C$. The functions $h_n$ will smoothly interpolate between
$\hs$ outside $F_C$ and $\hs$ just inside $F_{c_0}$ in such a way
that
$$
0 < \| \nabla h_n \| <_\textup{ess} O(n^{-1/2})\,, \quad n\to\infty \,.
$$
between $F_C$ and $F_{c_0}$. Here $<_\textup{ess}$ means that
the inequality is satisfied except on a set whose measure
goes to zero as $n\to\infty$.

The derivative of $\cE$ at $h_n$
in the direction $g-h_n$ equals
$$
\int (c - \dv \nabla \sigma (\nabla h_n)) \, (g - h_n) \,,
$$
which is nonnegative for all sufficiently large $n$ provided
$g > \hs$ somewhere on the facet and
$$
\dv \nabla \sigma (\nabla h_n) \le c
$$
between $F_C$ and $F_{c_0}$.

The functions $h_n$ will be obtained by patching
together solutions of
\begin{equation}
  \label{ELi}
    \dv \nabla \sigma (\nabla f_i) = c_i \,,
\end{equation}
where
$$
c =c_0> c_1 > c_2 > c_3 > \dots > c_n=C\,,
$$
are chosen so that all the distances between the
fronts $F_{c_i}$ go to zero as $O(n^{-1})$ as
$n\to\infty$. The implicit constants in $O(\dots)$ here and
below are bounded in terms of $C$.

For each $c_i$, consider the algebraic solution $f_i$ of \eqref{ELi}
with frozen front $F_{c_i}$. Since near the frozen boundary
its gradient has a square root growth, we have
$\|\nabla f_i \|=O(n^{-1/2})$ between $F_{c_i}$ and
$F_{c_{i-1}}$. This means that the discontinuous surface
formed by the union of $f_i$'s between $F_{c_i}$ and
$F_{c_{i-1}}$ is nearly flat
for $n\gg 0$. To make it continuous, we shift each successive
piece slightly in the $x_3$ direction (recall that our
facet is assumed to be horizontal) and insert a smooth
nearly vertical
strip of surface between two $f_i$'s. The width of each of these strips is
$O(n^{-3/2})$ and so their contribution to the functional
is negligible. There will be also a small region near the
boundary of $\Omega$ where $f_i$ is not defined. Since
each $F_{c_i}$ is tangent to $\partial\Omega$, the area
of that region is $O(n^{-3})$ and so we can extend $f_i$ there
as an arbitrary smooth Lipschitz function. Thus we have
constructed the required sequence $\{h_n\}$.
\end{proof}

\section{Explicit examples}

In this section
we work out some explicit solutions to \eqref{PQ} for various cases.

It can be checked along the lines of Section \ref{sgmin}
that the solutions we construct below
are minimizers for their corresponding variational
problem.

\subsection{Disconnected boundary example}\label{nonsimplyconnsec}

The simplest multiply-connected regions are obtained from
simply-connected regions by removing points.  In this case the
inner boundary components are reduced to points and
specifying a surface with these boundary conditions is
equivalent to specifying the height of the surface at
given points in the interior of the domain.

\begin{figure}[!htbp]
  \centering
\vspace{-0.5 cm}
\rotatebox{30}{\scalebox{.5}{\includegraphics{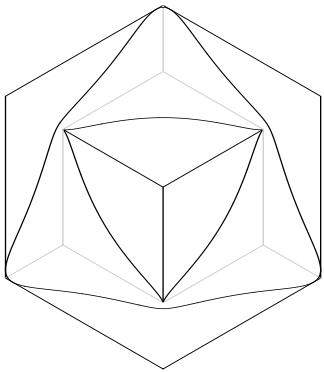}}}

\vspace{-1.5 cm}
  \caption{Boxed plane partition conditioned to pass through a given point.
The gray edges lie below the surface.
  \label{bppb}}
\end{figure}

Figure \ref{bppb} shows an example:
here the outer contour is a hexagon ($6$ edges of a cube)
and the inner contour
is a single point, the central point in the figure,
located at a height which is $2/3$ of the way
between the upper and lower vertices of the corresponding cube.

The curve $Q$ is a cubic curve which by symmetry
has the form
$$
Q(z,w)=\begin{array}{llll} w^3\\+bw^2&+bzw^2\\+bw&+azw&+bz^2w
\\+1&+bz&+bz^2&+z^3.
\end{array}
$$
The coefficient $b$ is determined by the intersection of $Q$ with
coordinate axes, that is, it determined by the size of the cube and
the value of Lagrange multiplier $c$. The coefficient $a$ determines
the height of the central point.

\subsection{Example with a bubble}\label{bubbleex}

\begin{figure}[!htbp]
  \centering
  \scalebox{0.5}{\includegraphics{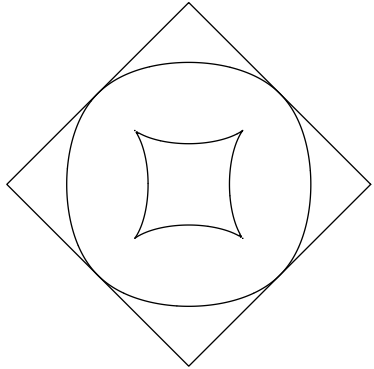}}
  \caption{frozen boundaries for the square-octagon ``fortress" example.
  \label{fortpic}}
\end{figure}

In this section we consider dimers on the square-octagon graph.
Here $P(z,w)=5+z+1/z+w+1/w$, see \cite{KOS}.
The Newton polygon is defined by $|x|+|y|\leq 1$.
Consider the contour $C$ through the points
$$
(1,0,0),(0,1,1),(-1,0,0),(0,-1,1)\,,
$$
see Figure \ref{fortpic}.

We expect $4$ frozen phases and hence $4$ points where the 
frozen phases change along the frozen boundary. Such points 
correspond to points of $P$ at the toric infinity, that is, tips
of tentacles of the amoeba of $P$. We, therefore, expect 
$Q$ to have the same Newton polygon as $P$. From symmetry, 
we conclude that 
$$
Q(z,w)=b+z+1/z+a(w+1/w)\,.
$$
The point $(z,w)=(\infty,\infty)$ on $P$ occurs
somewhere along the upper right boundary edge (the line $x+y=1$)
since this is the point of separation
of the two adjacent frozen phases. Eliminating $w$ using \eqref{PQ},
substituting $y=1-x$, and expanding near $z=\infty$
gives 
$$
a=e^{-c+2cx_0}\,, \quad b=5e^{-2c+3cx_0}\,,
$$
where $(x_0,1-x_0)$ is the point of intersection of the frozen boundary with
the line $x+y=1$. This leaves us with one variable, $x_0$,
which is determined by the area constraint from 
Proposition \ref{log_arias}.

While it is a non-trivial computation to compute $x_0$ in general, in the
case $c=0$ we have an extra symmetry and expect
$x=1/2$. This leads to a degree $8$ equation
\begin{multline}\label{octc}
729 - 13608x^2 - 22896x^4 + 64000x^6 + 102400x^8 -
  13608y^2 + \\412992x^2y^2 - 1104000x^4y^2 +
  870400x^6y^2 - 22896y^4 - 1104000x^2y^4+ \\
  2054400x^4y^4 + 64000y^6 + 870400x^2y^6 +
  102400y^8=0
\end{multline}
for the frozen boundary in the unconstrained volume case--see Figure \ref{fortpic}.
This was called the ``octic circle'' by Cohn and Pemantle \cite{CP}.

This case can be obtained more simply using (\ref{Q when c=0}):
there let $Q_0(z,w)\equiv -3/2$.
We solve
\begin{eqnarray*}
     P(z,w)&=&0\\
  -\frac{3}2&=&xzP_z+ywP_w
 \end{eqnarray*}
 for $z,w$ as a function of $x,y$. The double root of these
equations occurs along \eqref{octc}, as one immediately checks
using resultants.

\subsection{Crystal corner boundary conditions}\label{minksum}

In this section we consider minimizers defined on
the whole plane and satisfying piece-wise linear
boundary conditions at infinity. They can be
interpreted as dissolution shapes of an
infinitely large crystal near one of its corners.

\subsubsection{Minkowski sums of Ronkin functions}

Recall that the Minkowski sum of two sets $A,B\in\R^n$
is $$A+B=\{x+y~|~x\in A,~~y\in B\}.$$
If $A$ and $B$ are convex, so is their sum.

For a curve $Q(z,w)$ let $A_Q$ be the set of points in $\R^3$
lying on or above the graph of the Ronkin function $\Ronk_Q$ of $Q$.
This is a convex set.

A general class of solutions to the volume-constrained minimization
problem for an arbitrary spectral curve $P$ can be constructed as follows.

\begin{Theorem}\label{mink}
Let $P'$ be a Harnack curve with the same Newton polygon as $P$
such that the areas of the corresponding (compact) facets of $\Ronk_P$ and 
$\Ronk_{P'}$ are equal. Then the boundary of the Minkowski sum of the sets
$A_P$ and $A_{P'}$ is a minimizer.
\end{Theorem}

\begin{proof}
It is a simple geometric property of the Minkowski sum that
a facet on the boundary of $A_P+A_{P'}$
is the Minkowski sum of a facet of $A_P$ and a facet of $A_{P'}$ having the
same slopes,
where one of these facets may be reduced to a point.
Any point $p$ on the boundary of $A_P+A_{P'}$ but not on a facet 
is the sum of unique points on the boundary of $A_P$ and $A_{P'}$ with
tangent planes in the same direction.

Given a point $(x_1,y_1,\Ronk_P(x_1,y_1))$ on the graph of $\Ronk_P$, which is
not in the interior of a facet of $\Ronk_P$, there is a corresponding
point $(z_1,w_1)$ on the curve $P$, with $\log|z_1|=x_1$ and $\log|w_1|=y_1$,
and such that the
arguments of $z_1$ and $w_1$ are linearly related to
the slope of $\Ronk_P$ at $(x_1,y_1)$.
The point $(x_2,y_2,\Ronk_{P'}(x_2,y_2))$
with the same slope on the graph of $\Ronk_{P'}$ (and if this point is
on the boundary of a facet, it must also have the same slope along the facet
boundary as the corresponding point of $\Ronk_P$) yields a
corresponding point $(z_2,w_2)$ lying
on the curve $P'$ and
with the same arguments.

Let $Q(z,w)=P'(1/z,1/w)$. Construct the minimizer using the curve
$Q$ in \eqref{PQ}.
Since $Q(e^{-cx}z_1,e^{-cy}w_1)=0$, we have
$(z_2,w_2) = (e^{cx}/z_1,e^{cy}/w_1)$.
Therefore
\begin{multline*}
  (x_2,y_2) = (\log|z_2|,\log|w_2|)=\\(cx-\log|z_1|,cy-\log|w_1|)
=c\cdot(x,y)-(x_1,y_1)\,,
\end{multline*}
and so $(x,y)=\frac1{c}((x_1,y_1)+(x_2,y_2))$.

The scaled Minkowski sum is the unique shape whose tangent plane above $(cx,cy)$
has the same slope as the tangent planes at $(x_1,y_1)$ and $(x_2,y_2)$. Therefore
it is the solution to the minimization problem \eqref{PQ}, except that 
it may not satisfy the area constraints of Proposition \ref{log_arias}. 
Those say precisely that the facet areas are equal. 
\end{proof}

In \cite{KO} we proved that, for fixed boundary behavior, there is a unique
Harnack curve with facets of given areas.
So we can always find $P'$ with the appropriate facet areas.

Note that if the facet areas of $P$ and $P'$ do not match,
we still get a  $C^{3/2}$ function that solves the Euler-Lagrange
equation in the liquid region and is locally a minimizer inside
each facet. It is not a global minimizer since
raising or lowering a facet as a whole will reduce the
total surface tension. {F}rom the point of view of Glauber
dynamics on random surfaces, moving the facet as a whole
is an extremely slow process, so this solution of the
Euler-Lagrange equation may be described as \emph{metastable}.

Note also the symmetry between $P$ and $P'$: $A_P+A_{P'}$ is a minimizer
for the variational problems for both $P$ and $P'$.

There exist intriguing connections, first noticed in
\cite{ORV}, between the random surfaces
studied in this paper and topological sting theory, see e.g.\
\cite{O} for a review aimed at mathematicians.
In this context, the Minkowski sum construction has
a natural interpretation which will be explained in \cite{KOV}.

\subsubsection{Wulff construction}

Taking $P'=P$ in Theorem \ref{mink},
we have $A_P=A_{P'}$ and the
facets trivially have the same area. So
limit shape is a copy of the Ronkin function of $P$.
This shows that the Ronkin function itself is a solution to the
volume-constrained surface-tension minimization problem.
Indeed, this is the classical Wulff construction, since the Ronkin function
is the Legendre dual of the surface tension \cite{KOS}.

\subsubsection{Higher-degree solutions}

To combine $P$ with higher-degree curves in Theorem \ref{mink},
one can take covers of $P$ as
described in \cite{KOS}: the curves
$$P_n=\prod_{\eta^n=\xi^n=1}P(\eta z^{1/n},\xi w^{1/n})$$
have the same Ronkin function up to scale as $P$,
and $N(P_n)=n N(P)$, where $N(P)$ stands for the 
Newton polygon of $P$. 

Let $Q$ be a Harnack curve with $N(Q)=N(P_n)$, and having
facets of the same area as $P_n$. Then $A_{P_n}+A_{Q}$ is a scaled minimizer
for $P_n$, and since $\sigma_{P_n}(ns,nt)=\sigma_P(s,t)$,
$A_{P_n}+A_Q$ is also a scaled minimizer for $P$.

This allows us to combine $P$ with arbitrarily
high-degree Harnack curves, and in fact by taking limits
one can construct explicit non-algebraic minimizers.

\subsubsection{Rational curves}

A particular case of Theorem \ref{mink} is when
$P$ has genus zero---then $Q$ must also have genus zero.
Suppose $P(z,w)=z+w+1$. Then $P_n$ can be parameterized
via
$z\mapsto(z^n:(-1-z)^n:1)$. A degree-$n$ rational Harnack curve $Q$ has
a parametrization
$$u\mapsto\left(
A\prod_{i=1}^n(u-a_i) : B\prod_{i=1}^n(u-b_i): C\prod_{i=1}^n(u-c_i)\right)$$
with $A,B,C\in\R$ and
where the real numbers $a_1,\dots,a_n,b_1,\dots,b_n,c_1,\dots,c_n$
occur in cyclic order around $\R\cup\infty$
(that is, all the $a$'s come first, then the
$b$'s and then the $c$'s), see \cite{KO}.

Then \eqref{PQ} gives
\begin{eqnarray*}
z^n&=&e^{cx}\frac{C}{A}\prod_{i=1}^n\frac{u-c_i}{u-a_i}\\
(-1-z)^n&=&e^{cy}\frac{C}{B}\prod_{i=1}^n\frac{u-c_i}{u-b_i}.
\end{eqnarray*}

The constants $c,A,B,C$ can be incorporated into an 
overall scale and translation. Suppose that as $n\to\infty$, 
the measures 
$$
\frac1{n}\sum \delta_{a_i}\,, \quad  \frac1{n}\sum \delta_{b_i}\,, \quad 
\frac1{n}\sum \delta_{c_i}
$$
have weak limits that we will denote by $d\mu_i$, $i=1,2,3$. 
Taking logarithms and limit, we find that the triangle 
discussed in Section \ref{s_recons} takes the form 
\begin{equation}
\sum_{i=1}^{3} e^{c_0 x_i+c_i} \exp\left(-\int\log(u-t)d\mu_i(t)\right) =  0\,,\end{equation}
for certain constants $c_0,\dots,c_3$.


\begin{thebibliography}{99}


\bibitem{Aba}
A.~Abanov,
\emph{Hydrodynamics of correlated systems.
Emptiness Formation Probability and Random Matrices},
Applications of random matrices in physics, 139--161, NATO Sci. Ser. II Math. Phys. Chem., \textbf{221}, Springer, Dordrecht, 2006. cond-mat/0504307.


\bibitem{CP} H.~Cohn, R.~Pemantle, unpublished, communicated to the authors.

\bibitem{CKP}
H.~Cohn, R.~Kenyon, J.~Propp,
\emph{A variational principle for domino tilings},
J.\ Amer.\ Math.\ Soc., {\bf 14}(2001), no.~2, 297-346.



\bibitem{GH}
Ph.~Griffiths and J.~Harris, 
\emph{Principles of Algebraic Geometry}, John Wiley \& Sons, 
New York, 1994. 

\bibitem{Gui}
A.~Guionnet,
\emph{First order asymptotics of matrix integrals;
a rigorous approach towards the understanding of matrix models},
  Comm.\ Math.\ Phys.\  \textbf{244}  (2004),  no. 3, 527--569.

\bibitem{Fournier}
J.-C.~Fournier, \emph{Pavage des figures planes
sans trous par des dominos: fondement graphique de l'algorithme de
Thurston et parall\'elisation}, Compte\ Rendus\ del L'Acad.\ des Sci.,
Serie~I \textbf{320} (1995), 107--112.


\bibitem{FP}
W.~Fulton and R.~Pandharipande,
\emph{Notes on stable maps and quantum cohomology},
Algebraic geometry---Santa Cruz 1995, 45--96,
Proc.\ Sympos.\ Pure Math., 62, Part 2,
AMS, Providence, RI, 1997.


\bibitem{HM}
J.~Harris and I.~Morrison, 
\emph{Moduli of curves}, Springer, 1998. 


\bibitem{ItVir}
I.~Itenberg and O.~Viro,
\emph{Patchworking algebraic curves disproves the Ragsdale conjecture},
  Math.\ Intelligencer  \textbf{18}  (1996),  no.~4, 19--28.


\bibitem{K.fluct}
R.~Kenyon,
\emph{Height fluctuations in honeycomb dimers}, math-ph/0405052.

\bibitem{KO}
R.~Kenyon, A.~Okounkov,
\emph{Dimers and Harnack curves},
Duke Math.~J. \textbf{131} (2006), no.~3, 499-524.

\bibitem{KOV}
R.~Kenyon, A.~Okounkov, C.~Vafa,
in preparation.


\bibitem{KOS}
R.~Kenyon, A.~Okounkov, S.~Sheffield,
\emph{Dimers and Amoebae},
Annals of Math. \textbf{163} (2006), no.3, 1019-1056.


\bibitem{Mat}
A.~Matytsin,
\emph{
On the large-$N$ limit of the Itzykson-Zuber integral}
Nuclear Phys.\ B \textbf{411} (1994), no.~2-3, 805--820.

\bibitem{Mikh}
G.~Mikhalkin,
\emph{Amoebas of algebraic varieties and tropical geometry},
Different faces of geometry,  257--300, Int.~Math.~Ser.,
Kluwer/Plenum, New York, 2004, math.AG/0403015.

\bibitem{Mikh2}
G.~Mikhalkin,
\emph{Enumerative tropical algebraic geometry in $\R^2$},
J. Amer. Math. Soc. \textbf{18}  (2005), no.~2, 313--377

\bibitem{Morrey}
Ch.~Morrey, Jr.,
\emph{Multiple integrals in the calculus of variations},
 Die Grundlehren der mathematischen Wissenschaften,
Band 130, Springer-Verlag New York, Inc., New York 1966.


\bibitem{O}
A.~Okounkov,
\emph{Random surfaces enumerating algebraic curves},
European Congress of Mathematics, 751--768, Eur. Math. Soc., ZŸrich, 2005.


\bibitem{ORV}
 A.~Okounkov, N.~Reshetikhin, C.~Vafa,
\emph{Quantum Calabi-Yau and Classical Crystals},
The unity of mathematics, 597--618, Progr. Math., \textbf{244}, BirkhŠuser Boston, Boston, MA, 2006.

\bibitem{PR}
M.~Passare and H.~Rullg\aa rd,
\emph{Amoebas, Monge-Ampère measures, and triangulations
of the Newton polytope},
Duke Math.~J.\  \textbf{121}  (2004),  no.~3, 481--507.

\bibitem{PT}
V.~Pokrovsky, A.~Talapov,
\emph{Theory of two-dimensional incommensurate crystals},
JETP (Zhurnal Experimentalnoi i Teoreticheskoi Fiziki),
\textbf{78} (1980), no.~1, 269-295.

\bibitem{Spey}
D.~Speyer,
\emph{Horn's Problem, Vinnikov Curves and the Hive Cone},
Duke Math.~J. \textbf{127} (2005), no.~1, 395-427.


\bibitem{Vin}
V.~Vinnikov,
\emph{Selfadjoint determinantal representations of real plane curves},
Math.\ Ann.\ \textbf{296} (1993), no.~3, 453--479.



\end{thebibliography}
\end{document}